\documentclass{aa}  

\usepackage{graphicx}
\usepackage{txfonts}
\usepackage{amsmath,amsfonts}
\usepackage{subcaption}         
\usepackage{lscape}             
\usepackage{placeins}

\usepackage{xcolor}

\begin{document}

   \title{The rotation-magnetism relationship in solar-type stars} 
   \subtitle{Constraining magnetic flux emergence rates}
   \author{Emre I\c{s}{\i}k\inst{1} 
   \and Sami K. Solanki\inst{1,2}
   \and Natalie A. Krivova\inst{1}
   \and Alexander I. Shapiro\inst{3,1}
        }
   \institute{Max-Planck-Institut f\"ur Sonnensystemforschung, 
   Justus-von-Liebig-Weg 3, 37077 G\"ottingen, Germany \\
             \email{isik@mps.mpg.de; e.ishik@gmail.com} \and 
    School of Space Research, Kyung Hee University, Yongin, Gyeonggi 17104, Republic of Korea \and
    Institut f\"ur Physik, Universit\"at Graz, Universit\"atsplatz 5/II, 
   8010 Graz, Austria
             }

   \date{}

  \abstract
   {The rotation-activity relationship of G-type stars results from 
   surface magnetic fields emerging from the interior. How the magnetic 
   flux and its emergence rate scale with rotation rate are not well understood, 
   both observationally and theoretically.} 
   {We aim to constrain the emerging magnetic flux as a function of the 
   rotation rate in solar-type stars by numerical simulations compared to empirical 
   constraints set by direct measurements of stellar magnetic fields.} 
   {We used our flux emergence and transport (FEAT) model for stars with a range 
   of power-law 
   slopes for the dependence of the emerging flux on rotation. Complementing this 
   with a heuristic account of the main flux components, we modelled 
   the resulting mean unsigned field strength as a function of the rotation 
   rate. We compared the results with the Zeeman-intensification measurements and spectropolarimetric data of solar-type stars. }
   {Deviations of the model from observations of G stars correlate strongly with stellar metallicity 
   ($r=0.83$) and effective temperature ($r=-0.76$), with a combined coefficient 
   of 0.90, reflecting the dependence of magnetic activity on these 
   two parameters. Correcting for these effects with multilinear regression, 
   we find that magnetic flux emergence rates must scale steeply with rotation 
   (power-law exponent of about 1.9) to reproduce observed field strengths, 
   significantly exceeding the estimates in the literature. We also provide 
   correction factors for metallicity and temperature for measurements of 
   early-G-type stellar magnetic fields.}
   {Stellar magnetic flux emergence rates scale steeply with rotation, requiring active-region fields to dominate the total surface flux on rapid rotators, whereas small-scale-dynamo fields dominate for slow rotators such as the Sun. Metallicity significantly influences the rotation-magnetism relationship, necessitating sample-dependent corrections for accurate stellar dynamo modelling.}

   \maketitle
   \nolinenumbers

\section{Introduction}

Magnetic activity on cool stars with a given effective 
temperature is empirically well correlated with the rotation rate. 
This is evident in the so-called rotation–activity relations observed through several indicators from the photosphere to the corona \citep{SchrijverZwaan00,Guedel04,Hall08,Basri21}, and reflects the increasing magnetic flux that emerges through the stellar surface with faster rotation \citep{Solanki06,Reiners12,Weber23}. 
Solar-type (G2V) stars are expected to have similar surface magnetic structures with respect to the Sun but with spatial distributions that evolve  with age as stars spin down through magnetised winds \citep{Bouvier14,Isik23}.

Rotation-activity relations are well studied
from both observational \citep{Noyes84,Baliunas95,Radick98,Pizzolato03,Reiners14}
and theoretical \citep{Karak14,Blackman15,Farrish21} perspectives. 
Still, one key quantity that directly drives stellar activity~--
the unsigned surface 
magnetic flux~-- has been comparatively rarely studied \citep{Vidotto14}. 
Even more poorly constrained is the emergence rate of magnetic flux at the stellar surface. 
It has been studied mainly for the Sun \citep{HarveyZwaan93,ThorntonParnell11,Krivova21}, 
but remains largely unconstrained on other
solar-type stars \citep[see][for a recent example]{Jeffers22}.

To measure stellar surface magnetic fields, two main observational techniques have been employed. 
Spectropolarimetry \citep[Zeeman-Doppler imaging, ZDI;][]{Semel89,Donati+Brown97} allows one to map 
large-scale field geometry and detect magnetic cycles
\citep{Bellotti25a,AlvaradoGomez25}, but
suffers from cancellation of mixed-polarity small-scale flux 
within spatial resolution elements 
\citep[see][for estimated corrections]{See19}. 
Zeeman broadening of absorption lines, 
measured in Stokes I, provides the total unsigned 
photospheric field independent of polarity \citep{Robinson80,Saar86,Basri88,Ruedi97}. 
This method has been successfully applied to mid-K to late-M-type stars 
\citep{Reiners22}
but becomes less reliable for G-type stars due to 
significant Doppler broadening and a weaker mean magnetic field.
This limitation was recently overcome by \citet{Kochukhov20}, who used relative Zeeman intensification (desaturation) of magnetically sensitive 
Fe~I lines in the presence of stellar magnetic fields 
to measure the mean unsigned field strengths of 14 solar-type stars 
including the Sun as a star. 
Their results confirmed a strengthening of the magnetic field with 
an increasing rotation rate. 
However, Zeeman intensification measures the disc-integrated unsigned field without revealing which magnetic structures produce the signal.

On the Sun, 
the surface magnetic field has two main components.
$(1)$ Small-scale magnetic flux covers most of the quiet
photosphere and is continually generated
by a small-scale dynamo (SSD) 
driven by near-surface convection 
\citep{Voegler+msch07,Danilovic10a,Danilovic10b,Rempel14}.
The SSD flux is additionally
replenished by decaying active regions (ARs).
$(2)$ Active-region flux emerges in the form of bipolar magnetic regions (BMRs) and their surrounding facular fields, which dominate solar irradiance variability on solar rotation timescales and form bright chromospheric plage \citep[see][for a review of the effect of magnetic fields on irradiance]{Solanki13}.

The detailed structure of magnetic fields and their rotational dependence on other solar-type stars remains uncertain.
In this study, we address these using the flux emergence and transport (FEAT) model \citep{Isik18,Isik24}, which combines buoyant rise of magnetic flux tubes with their surface evolution, to estimate what rotation dependence of emerging flux is required to reproduce the observed mean unsigned field strengths. We used FEAT to model the AR-driven large-scale field, which is expected to respond most sensitively to rotation. In contrast, SSD fields are driven by near-surface convection and are not expected to change considerably with rotation rate \citep{Rempel23}. We therefore assume that the observed rise of the mean field towards faster rotators is primarily related to the rotation dependence of the emergence rate of flux generated by the global dynamo. In the original FEAT model \citep{Isik18}, this scaling was assumed to be linear with rotation.

We complemented FEAT simulations with empirically determined small-scale BMR flux that is not accounted for, and scaled the emergent flux with rotation on top of a constant SSD background. We then compared the model against magnetic field 
measurements of early G-stars rotating faster than the Sun, to constrain how the emerging magnetic flux 
should scale with rotation. When comparing the model results with observations, we 
also took into account the influence of stellar physical parameters, such as 
effective temperature and metallicity, which can introduce  
systematic effects and scatter in rotation-activity relations. 

In Sect.~\ref{sec:model}, we describe the FEAT model set-up and how we 
scaled the emerging magnetic flux under observational constraints. 
In Sect.~\ref{sec:results}, we compare FEAT results with the measured 
mean fields of solar-type stars and analyse the role of additional 
stellar parameters.
In Sect.~\ref{sec:discuss}, we discuss the implications of our analysis  
for stellar dynamos and suggest new observational strategies 
as well as model improvements. Section~\ref{sec:conc} summarises our 
conclusions.

\section{Modelling the rotation-magnetism relationship}
\label{sec:model}

To estimate how the mean magnetic field scales with rotation, 
we used the FEAT framework developed by
\citet{Isik18,Isik24}. It 
calculates the surface distribution of magnetic flux (and its variation 
in time) as a function of the rotation rate and the activity level. 
The FEAT framework consists of two modules. 

The `emergence' module generates 
a synthetic solar-cycle record~-- including longitudes, latitudes, 
and times of emergence~-- at the surface and maps it to the base 
of the convection zone, accounting for the solar differential 
rotation rate. The map at the base latitudes 
is then scaled back to the surface for the given rotation rate using 
pre-calculated flux-tube rise tables.
This yields the surface emergence latitudes and 
tilt angles of flux loops, both of which are sensitive to the rotation rate. 

The `transport' module (i.e. surface flux transport; hereafter SFT) takes the time, latitude, and tilt angle from the emergence module and produces a corresponding
BMR for each emergence event. 
The BMR areas are drawn randomly from the observed solar distribution 
of sunspot group areas, converted to AR areas \citep{Jiang11}. 
The distribution of the sunspot-group areas is a power law with 
index $-1.1$ for up to the threshold area of 300~millionths of 
hemispheric area, above which the distribution becomes log-normal. 
The cycle has a fixed period of 10.4~years, close to that of solar cycle~22. Since we focus only on the activity maximum phase (100~days around it), the precise cycle period does not affect our results.
The polar fields are reversed on 
a timescale determined by the emerging flux and intense polar fields of 
the new polarity form already close to the activity maximum 
\citep[see][for details]{Isik18}. 
Finally, we compare global averages of surface magnetic 
field strength to direct measurements of magnetic fields in G-type 
main-sequence stars (Sect.~\ref{ssec:rot-B}). 

In the transport module, the emergence times, latitudes, and tilt 
angles are used as inputs to the SFT 
model, which integrates the magnetic induction 
equation for the radial field in time, using spherical harmonics up to a 
maximum degree of $\ell=64$, to cover the supergranular convective scales 
on the Sun. The surface magnetic flux, which is fed by freshly 
emerging BMRs, is advected by prescribed differential rotation and 
meridional flow, while subject to diffusion that is parametrised by 
the turbulent magnetic diffusivity at the supergranular convective scales 
\citep[][and the references therein]{Isik18}. 

\subsection{Scaling the emerging magnetic flux with rotation}
\label{ssec:scaling}

To model the rotation dependence of magnetic flux emergence, we employed 
the thin–flux-tube rise calculations described in \citet{Isik18,Isik24}.
For each rotation rate, $\omega:=\Omega_\star/\Omega_\odot\in \{1,2,4,8\}$, 
we pre-computed
a grid of rising flux tubes from the base of the convection zone to 
a radius of about 0.98$R_\star$, where the thin-flux-tube approximation 
breaks and further tracking of the tube apex into the photosphere is no 
longer possible. The grid provides emergence latitudes and tilt angles 
used to map subsurface toroidal (here
used in the dynamo sense of the azimuthal component of the subsurface
field, following \citealt{Parker55}) flux into photospheric BMRs.
For $\omega\geq 4$, we followed the procedure developed by \citet{Isik24} 
to allow low-latitude emergence despite rapid rotation. 

As in \citet{Isik18},
we took the solar cycle 22 with the maximum monthly mean sunspot number of 
$s_\odot=156$ as the reference activity level.
The emerging BMR sequence is described by the total number of 
emergences during the 10-year cycle and the cycle strength 
relative to the solar cycle 22 reference.
We also assumed an AR nesting degree of $60\%$, similar to the 
solar value \citep{Karapinar26}. 

The total unsigned magnetic flux injected by emerging BMRs is controlled 
by the parameter $s$.
We prescribed its rotation dependence as a power law:
\begin{eqnarray}
    s &=& s_\odot \omega^p,
    \label{eq:s-rot}
\end{eqnarray}
where the unknown power-law index $p$ is constrained by comparison with 
stellar observations.
The value of $p$ determines how strongly the emerging flux increases 
with rotation.

The initial large-scale dipole-like field at the start of each activity 
cycle is scaled similarly,
\begin{equation}
B_{\rm pol,0} = B^\odot_{\rm pol,0}\omega^p, 
\label{eq:Bpol-rot}
\end{equation}
as in the original FEAT model.
We thus scale the initial polar field with the same exponent $p$, 
purely to keep the SFT model internally consistent. 
On the Sun, the amplitude of the polar field at the start of a cycle 
correlates with the strength of the subsequent cycle 
\citep[e.g.][]{Cameron10}. 
Because FEAT scales the cycle strength with rotation, we apply the 
same scaling to the initial polar field so that the model remains 
balanced. We make this scaling under the assumption that polar 
fields are reversed and amplified by the emergence and decay of 
systematically tilted BMRs, similar to the solar case (see Sect.~\ref{ssec:caveats} for further discussion). 

In the SFT model, emerged BMRs evolve subject to differential rotation, 
meridional flow and diffusion. The resulting surface-averaged unsigned 
field at activity maximum then scales approximately as 
    \begin{equation}
        \langle |B_{\rm SFT}|\rangle \propto \omega^q,
        \label{eq:B-rot}
    \end{equation}
where the exponent $q$ describes the net rotation dependence of the 
surface field. Here, we evaluate $\langle |B_{\rm SFT}|\rangle$ as the unsigned surface-averaged field, 
averaged over 100 days around the cycle maximum to represent 
sufficient number of stellar rotations regardless of the rotation 
rate. 

\begin{figure*}
    \sidecaption
    \includegraphics[width=12cm]{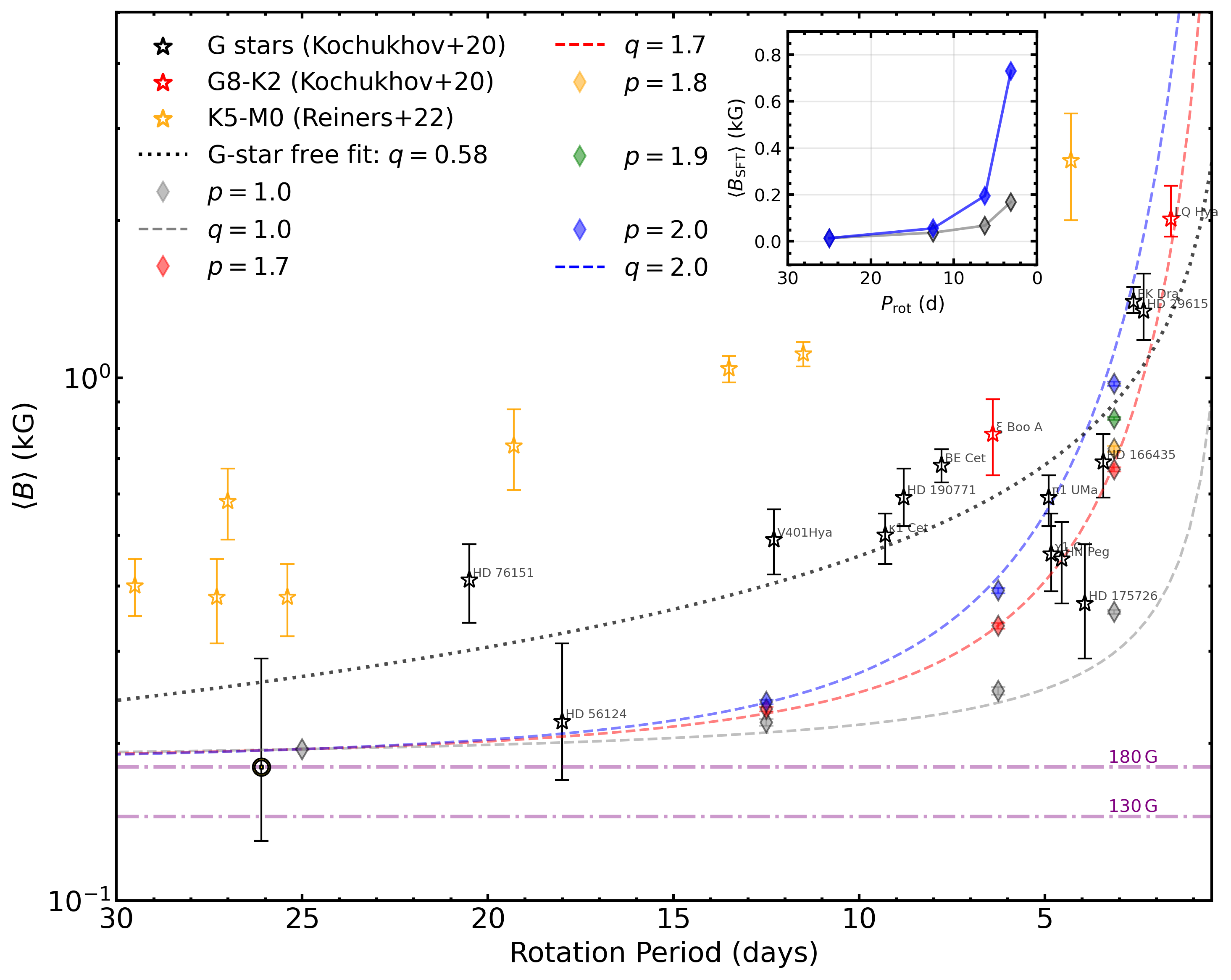}
    \caption{Rotation-rate dependence of the globally averaged unsigned field strength, 
    modelled with FEAT simulations for solar-type stars (coloured diamonds 
    with the flux injection exponent $p$ indicated in legend), scaling relations 
    based on the solar simulation at $P_{\rm rot}=25$~d (dashed curves with the 
    scaling exponent, $q$, indicated in the legend), in comparison to  
    direct measurements using Zeeman intensification by \citet{Kochukhov20} 
    for G stars (G8 and K2 types shown by red symbols) and 
    Zeeman broadening for 
    K5-M0 stars by \citet{Reiners22} -- see the legend in the figure. 
    The dotted curve shows a least-squares fit to the G-star data 
    (Sect.~\ref{ssec:scalings}). Error bars 
    denote the observed upper and lower margins for measurements and 
    the $1\sigma$ levels for the simulations averaged over 100 d. 
    The upper and lower horizontal lines denote the solar mean field 
    level as observed by 
    \citet{Kochukhov20} near the minimum and the mean SSD field strength adopted by \citet{Reiners22}, respectively. The inset plot shows the mean SFT field for $p=1$ and $p=2$.}
    \label{fig:rot-b}
\end{figure*}

In an absolutely linear SFT system without noise, $q$ would be equal to 
$p$. However, in practice, the emergence properties themselves depend 
on rotation: stronger Coriolis acceleration experienced by rising flux loops increases the tilt angles and 
latitudes of emerging BMRs, and high emergence rates lead to enhanced AR overlap and flux cancellation.
These effects cause $q$ to differ slightly from $p$, typically by no 
more than $\sim 0.1-0.2$ for the rotation rates considered here 
(Fig.~\ref{fig:rot-b}). Thus, $\langle |B_{\rm SFT}|\rangle$
is approximately proportional to the injected BMR flux, with small 
systematic deviations at the higher activity levels.

\subsection{Observational constraints}
\label{ssec:obs}

Measurements of Zeeman splitting of magnetic-sensitive lines from 
the visible to the infrared by \citet{Saar96} indicated a power-law 
dependence between the Rossby number ($Ro:=P_{\rm rot}/\tau_c$, 
the ratio of the rotation period to the convective turnover time) 
and the magnetic flux density. His result implies 
$\langle B\rangle_{\rm obs}\propto \omega^{1.7}$ ($q_{\rm obs}=1.7$) for 
solar-type stars.

An indirect estimate was found 
by combining the X-ray flux - rotation relationship $F_X \propto 
\Omega^{0.8\pm 0.2}$ with {\rm the X-ray - Ca II HK flux relationship, }
$F_X \propto \langle\Delta F_{\rm Ca II}\rangle^{1.5\pm 0.2}$ 
\citep{Schrijver92} 
and the relationship between the disc-integrated (solar) mean field 
strength and Ca II K flux, $\Delta F^*_{\rm Ca~II}\propto \langle 
B\rangle^{0.6}$ \citep{Schrijver+89}, 
resulting in $q_{\rm obs}=0.9\pm 0.3$ \citep{SchrijverZwaan00}. 
In parallel with this result \citep[see also][]{Reiners12}, 
the FEAT models presented in \citet{Isik18} assumed a 
linear scaling, by setting $p$ in Eq.~(\ref{eq:s-rot}) 
to unity. 

More recent studies have not led to 
a consensus on $q_{\rm obs}$ either. 
The Zeeman-intensification study of solar-type stars central to our analysis \citep[][K20]{Kochukhov20} gives a $Ro$-dependence with $q_{\rm obs}=0.67$ (assuming $\tau_c$ is constant across the sample, this is equivalent to $\langle B\rangle\propto\omega^{0.67}$). In their Zeeman broadening survey of GKM stars, where G-star data were adopted from K20, \citet{Reiners22} find $q_{\rm obs}=1.25$ for the unsaturated branch in the $Ro$ dependence. Assuming $\tau_c\propto L_{\rm bol}^{-1/2}$, this is equivalent to $\Phi_B\propto P_{\rm rot}^{-1.25}$, a relation that is independent of $\tau_c$.

The mean solar magnetic field is an important constraint to base our 
reference model and compare with measurements from other solar-type stars.
\citet{Kochukhov20} measured the mean field of the mostly quiet solar 
magnetic field, using the HARPS Sun-as-a-star spectrum taken in
November 2010, corresponding to the early ascending phase of the relatively 
weak solar cycle 24. By fitting synthetic profiles to one Zeeman-insensitive and three relatively strong and Zeeman-sensitive Fe~I lines, a 2D grid search between the magnetic field 
strength $B$ and its filling factor $f$, yielded $B=2.6^{+3.2}_{-1.2}$~kG 
and $f=0.07^{+0.07}_{-0.03}$, leading to a mean field strength of 
$\langle B_\odot\rangle_{\rm K20}=0.18^{+0.11}_{-0.05}$~kG. 
This mean field is likely dominated by the small-scale 
dynamo (SSD) fields manifested as the internetwork pattern 
and partly by the network fields processed in supergranular boundaries. 
We note also that in the observation epoch (2010-11-25T06:00:00), 
a diffuse field from one decaying AR was also present. 
$\langle B_\odot\rangle_{\rm K20}$ is considerably higher than
the measurement by \citet{TrujilloBueno04} of 130~G, though the latter is 
marginally within the lower error bound of the former value. Other 
measurements in the literature are 185~G \citep{Lites08}, 220~G 
\citep{OrozcoBellot12}, 130~G \citep{Danilovic16}, though 
all these values are likely upper limits, owing to selection effects 
\citep{Rempel23}. We adopt 
$\langle B_{\rm SSD}\rangle\equiv\langle B_\odot\rangle_{\rm K20} = 180$~G 
as the base SSD field 
and build up the rotation-dependent larger-scale field components additionally. The motivation for this choice 
is that \citet{Kochukhov20} used the same method and spectral lines to measure $\langle B\rangle$ 
of the other target stars in their sample, which we use here to compare with our models. 

\subsection{Building blocks of the mean surface magnetic field}
\label{ssec:rot-B}

Naturally, with its coarse grid ($360^\circ\times180^\circ$) the FEAT model does not resolve the intense 
fields that form the quiet photospheric network, nor the SSD (internetwork) 
fields: 
the field that the surface flux transport (SFT) module resolves, 
$\langle B_{\rm SFT}\rangle$, is only on the order of 10~G during the maximum of 
the synthetic cycle 22, involving the transported flux sourced from emerging 
ARs hosting sunspots. In addition, the SFT model does not resolve AR fields below a size of $1^\circ$ and particularly in small-scale ephemeral regions, which are all structured on very small scales. The model provides a spatially averaged field that can be scaled to match the resolution of solar magnetograms (see below).

To account for the missing flux in the FEAT model, we include the small-scale 
magnetic flux contributions in a heuristic fashion, in two components. 
The first one is the constant contribution mainly set by the SSD field, 
discussed in Sect.~\ref{ssec:obs}. The second one is related to small-scale 
BMRs emerging throughout the solar cycle, called small-scale emergences (hereafter 
SSEs), i.e. bipoles without sunspots, including the lowest-flux `ephemeral 
regions' \citep{Harvey73}, which bring far more flux to the solar surface than the large ARs. 

The remaining magnetic flux is composed of larger BMRs that form sunspots when they emerge. The transport module of FEAT includes mainly this component, where SSEs are only partially taken into account in the flux budget, as described in the following.
Each BMR is defined as a doublet of localised radial magnetic field of opposite signs. The field distribution in each unipolar magnetic patch is of the form
\begin{equation}
    B^{\pm}(\lambda,\phi)=B^\pm_{\rm max}\left(\frac{0.4\Delta\beta}{\delta}\right)^2 \exp\left[\frac{2\left(1-\cos\left[\beta_\pm(\lambda,\phi)\right]\right)}{\delta^2}\right],
\label{eq:BMR}
\end{equation}
where $\Delta\beta$ is the angular separation between opposite-polarity centres, $\delta=4^\circ$ is the intrinsic width of each polarity patch, and $\beta_\pm$ is the heliocentric angle between a given pair of latitude and longitude $(\lambda,\phi)$ and the polarity centres $(\lambda_\pm,\phi_\pm)$. The size and flux of the BMR are controlled solely by the separation $\Delta\beta$, which is connected to the AR area via $\Delta\beta = 0.45A_R^{1/2}$, empirically obtained by \citet{Cameron10}, 
who also determined an empirical BMR-field normalisation. This was done by setting 
$B^\pm_{\rm max}=\pm 374$~G, which scales the SFT model to match the 
observed strength of the disc-integrated line-of-sight field throughout solar cycles, measured 
at the resolution reached by Mount Wilson and Wilcox Solar Observatory magnetograms. 
This ensures that the model partly compensates for the flux in SSEs which are not directly included into our input BMR record based on sunspot observations. 

However, these low-resolution solar magnetograms themselves miss a large fraction of SSEs.
Indeed, \citet{Hofer24} could reproduce such magnetograms by including 
all of the flux from spot-bearing ARs but only about 40\% of 
the flux from SSEs.
The remaining $\sim$60\% of the SSE flux is simply not detected at that 
resolution \citep[see][]{krivova-solanki-2004}.
To avoid underestimating the unsigned magnetic flux, we therefore include 
this missing fraction explicitly when computing the stellar surface field.
The large-scale mean field in
all emergent regions, $\langle B_{\rm em}\rangle$, will thus be higher than 
in the low-resolution magnetograms by about $0.6B_{\rm SSE}$, where we adopt $B_{\rm SSE\odot}=1.64$~G for the 
solar activity maximum, consistent with \citet[][Fig.~4a-b]
{Hofer24}. 

The total unsigned mean field as measured by 
\citet{Kochukhov20} is then composed of the following components, 
ordered from the smallest to the largest spatial scales: the SSD, 
SSE, and the BMR-driven fields subject to SFT, leading to
\begin{eqnarray}
    \langle B_{\rm tot}\rangle &=& \langle B_{\rm SSD}\rangle + 
    0.6\langle B_{\rm SSE\odot}\rangle\omega^p + \langle B_{\rm SFT}(p)\rangle,
    \label{eq:Btot}
\end{eqnarray}
where angular brackets denote averaging over the sphere and time, and the SSE flux coverage is partitioned between the second and third terms.
The first term on the right-hand side denotes SSD and network fields, 
assumed to be independent of the rotation rate.
This is a simplifying assumption observationally motivated by measurable magnetic fields in slowly rotating stars \citep{Reiners22}. While simulations show that SSD properties depend on parameter regime \citep[e.g. magnetic Prandtl number;][]{Rempel23}, there are currently no definitive constraints on rotation-dependence of SSD saturation in solar-type stars.
The rotational scaling of the third term, i.e. field sourced by emerging BMRs is already inherent in the last term, with the dependence 
controlled by the flux-injection power index $p$ in Eq.~(\ref{eq:s-rot}) and 
the expected rotation dependence of the evolving large-scale field as in Eq.~(\ref{eq:B-rot}). 
The second term represents the majority of SSEs, which on the Sun show a weaker dependence on the sunspot cycle than the sunspot bearing ARs \citep{HarveyZwaan93}. 
For stars with significantly higher magnetic activity, the scaling of SSEs with rotation is not observationally constrained. 
Here we assume that the total unsigned flux associated with SSEs scales with the same power-law exponent $p$ as for BMRs. 
This is motivated by the fact that the number and latitude of SSEs changes over the solar cycle similar to the larger ARs, suggesting that they are also produced by the global dynamo. On more active stars, 
we therefore treat the $\omega^p$ scaling of the SSE contribution as a first-order approximation. 

To complement the discrete simulations, we define analytical 
reference curves $\langle B_{\rm tot}\rangle_q$ that take the solar-simulation 
field components as their anchor and scale the rotation-dependent part by 
$\omega^q$, while keeping the SSD background constant:
\begin{equation}
    \langle B_{\rm tot}\rangle_q = \langle B_{\rm SSD}\rangle + 
    \left[0.6\langle B_{\rm SSE\odot}\rangle + \langle B_{\rm SFT\odot}\rangle\right]\omega^q.
    \label{eq:Btotcurves}
\end{equation}
These curves serve two purposes: $(i)$ validating that the discrete 
simulations at $\omega\in\{1,2,4,8\}$ follow the expected power-law 
scaling (i.e.\ confirming $q\simeq p$), and $(ii)$ providing a smooth, 
continuous reference from which to evaluate observational residuals 
$\Delta B$ at arbitrary rotation rates (Sect.~\ref{ssec:correlation}).

Motivated by the empirical indications given in Sect.~\ref{ssec:obs}, we limit the simulations to the interval $1<p<2$. The expected scaling of the total 
solar field to faster rotators is then approximated by 
the extrapolation in Eq.~(\ref{eq:Btotcurves}), with 
$1<q<2$.

\subsection{Model assumptions and parameter sensitivity}
\label{ssec:assumptions}

Our model framework fixes several parameters to their solar values, which affects the absolute magnitude of predicted surface-averaged fields. However, the value of the power-law index $p$ required to match observations is robust to moderate variations in these choices, provided parameters do not themselves vary systematically with rotation rate. Test calculations with modified bipole separation distributions indicate that when absolute modelled field strengths change, the observational constraint on $p$ remains essentially unchanged, as these modifications affect all rotation rates similarly. The most critical fixed parameter is $B_{\rm amp} = 374$~G, which sets the normalisation of BMR field strengths; varying this shifts all model curves vertically without affecting the best-fit power-law index. The primary uncertainty lies in the potential rotation dependence of intrinsic BMR field strengths, which cannot presently be disentangled from emergence rate variations using surface-averaged field measurements.

\section{Comparisons with observed fields}
\label{sec:results}

Following the model set-up described in Sect.~\ref{sec:model}, we compare the surface-averaged unsigned field from the FEAT models, 
computed from Eqs.~(\ref{eq:Btot})--(\ref{eq:Btotcurves}),
with the surface-averaged magnetic field measurements of a subsample 
of early G-type stars studied by \citet[][K20 sample]{Kochukhov20}. 
In the present analysis, we excluded two stars from the K20 sample,
\object{LQ~Hya} (K1V, $T_{\rm eff} = 4936$~K) and
\object{$\xi$~Bo\"o~A} (G7V, $T_{\rm eff} = 5533$~K). As is evident
from Table~\ref{tab:kochukhov_sample}, the remaining 13 stars cluster
tightly in the range 5710--6080~K; these two stars lie well outside
that range and are clear $T_{\rm eff}$ outliers relative to the bulk
of the sample, making the application of a linear correction derived from that
bulk unreliable. Both stars are retained in all figures for reference
(shown with red symbols). In the Rossby-number comparison
(Fig.~\ref{fig:rossby}), \object{LQ~Hya} falls in the saturated activity
regime (${\rm Ro} = 0.067$), which our model does not address.
We do not associate \object{$\xi$~Bo\"o~A} (${\rm Ro} = 0.400$) 
with either the early-G-star or late-K to early-M samples, owing to its 
intermediate temperature making it a late-G-type star. Nevertheless, it 
sits in the unsaturated branch and is broadly consistent with the model 
curves and fit to early-G stars, as much as stars in the other samples.
The resulting sample will be called the G-star sample in the following.

\begin{table*}
\centering
\caption{Stellar parameters for the K20 \citep{Kochukhov20} sample.}
\label{tab:kochukhov_sample}
\begin{tabular}{llccccrc}
\hline\hline
HD & Name & Sp & $T_\mathrm{eff}$ & [Fe/H] & $P_\mathrm{rot}$ & Ro & $\langle B \rangle$ \\
   &      &    & (K)              &        & (d)               &    & (kG) \\
\hline
  1835 & BE\,Cet          & G3V   & 5787 & $+0.18$ &  7.78 & 0.659 & $0.68^{+0.05}_{-0.05}$ \\
 20630 & $\kappa^1$\,Cet  & G5V   & 5707 & $+0.04$ &  9.30 & 0.696 & $0.50^{+0.05}_{-0.06}$ \\
 29615 & ---              & G3V   & 5866 & ---     &  2.34 & 0.207 & $1.34^{+0.24}_{-0.16}$ \\
 39587 & $\chi^1$\,Ori    & G0V   & 5942 & $-0.04$ &  4.83 & 0.437 & $0.46^{+0.09}_{-0.07}$ \\
 56124 & ---              & G0V   & 5846 & $+0.01$ & 18.0  & 1.549 & $0.22^{+0.09}_{-0.05}$ \\
 72905 & $\pi^1$\,UMa     & G1.5V & 5888 & $-0.06$ &  4.90 & 0.437 & $0.59^{+0.06}_{-0.07}$ \\
 73350 & V401\,Hya        & G5V   & 5816 & $+0.11$ & 12.3  & 0.993 & $0.49^{+0.07}_{-0.07}$ \\
 76151 & ---              & G3V   & 5782 & $+0.10$ & 20.5  & 1.629 & $0.41^{+0.07}_{-0.07}$ \\
129333 & EK\,Dra          & G1.5V & 5755 & $+0.02$ &  2.61 & 0.223 & $1.40^{+0.09}_{-0.07}$ \\
166435 & ---              & G1IV  & 5804 & $-0.01$ &  3.43 & 0.293 & $0.69^{+0.09}_{-0.10}$ \\
175726 & ---              & G0V   & 6081 & $-0.04$ &  3.92 & 0.434 & $0.37^{+0.11}_{-0.08}$ \\
190771 & ---              & G2V   & 5796 & $+0.14$ &  8.80 & 0.742 & $0.59^{+0.08}_{-0.07}$ \\
206860 & HN\,Peg          & G0V   & 5933 & $-0.08$ &  4.55 & 0.481 & $0.45^{+0.08}_{-0.08}$ \\
\hline
131156A & $\xi$\,Boo\,A$^\dagger$ & G7V & 5533 & $-0.20$ & 6.40 & 0.400 & $0.78^{+0.13}_{-0.13}$ \\
 82558 & LQ\,Hya$^\dagger$ & K1V  & 4936 & $+0.17$ &  1.60 & 0.067 & $2.01^{+0.32}_{-0.15}$ \\
\hline
\end{tabular}
\tablefoot{ 
$T_\mathrm{eff}$ and [Fe/H] are from the PASTEL catalogue
\citep{PASTEL_Soubiran16}. Rotation periods $P_\mathrm{rot}$ and Rossby
numbers Ro are adopted from K20. Mean magnetic field
strengths $\langle B \rangle$ are measurements by K20. Stars
marked with $\dagger$ are shown in the figures but excluded from the
regression analysis (Sect.~\ref{sec:results}).
No [Fe/H] measurement is available in the literature for
HD\,29615; this star is shown in the figures but excluded from the
metallicity correlation analysis.}
\end{table*}

The K20 sample was assembled according to three explicit criteria
\citep[][Sect.~2.5]{Kochukhov20}: ($i$) fundamental stellar parameters
close to solar values, ($ii$) availability of prior ZDI observations, and
($iii$) availability of high-quality optical spectra suitable for Zeeman
intensification measurements. The first criterion is of particular
relevance to the present work: since all surface flux transport parameters
in the FEAT framework are calibrated to solar values, the near-solar
selection of K20 closely coincides with the regime in which our model can
be applied with confidence. The stellar parameters of the full K20 sample
are listed in Table~\ref{tab:kochukhov_sample}, where $T_\mathrm{eff}$ and
[Fe/H] are adopted from the PASTEL catalogue \citep{PASTEL_Soubiran16}. The
G-star sample spans effective temperatures of 5710--6080~K and rotation
periods of approximately 2--20~d. We note that our comparison with
observations is not limited to the K20 sample: in Sect.~\ref{ssec:zdi} we additionally
consider ZDI-based large-scale field measurements from \citet{See25}, and in
Sect.~\ref{ssec:imply} we consider the Rossby number dependence, including late-K star measurements from \citet{Reiners22}, finding good
agreement with the model predictions across this broader spectral type
range.

The results are presented in 
Fig.~\ref{fig:rot-b}, along with a set of late-K to early-M stars 
($T_{\rm eff}>3800$~K; hereafter the K5-M0 sample) from the
Zeeman-broadening measurements by \citet{Reiners22}. 
The K5-M0 stars follow a track roughly parallel to that of the G stars, but 
shifted to higher field strength, even at the solar rotation period, as 
expected from their smaller Rossby numbers at a given rotation period. 

The magnetic flux associated with the SSD is expected to be largely 
independent of the activity level.
Thus, the apparent strengthening of the measured fields 
with the rotation rate is most likely due to facular, and to some extent 
penumbral, fields associated 
with the greater number of ARs on more rapidly rotating stars. 
AR sizes may also vary, but for simplicity we assume that the increase in 
magnetic flux is solely controlled by the number of emergence events. We 
note that magnetic fields in G-star spot umbrae are likely poorly captured 
by Zeeman diagnostics in atomic lines, and penumbral fields are probably 
also somewhat underestimated.
As a result, the measured unsigned fields of active stars may 
systematically underestimate the true contribution of spot-bearing 
regions (see Sect.~\ref{sec:discuss}).

\subsection{Comparison of scaling relations}
\label{ssec:scalings}

Figure~\ref{fig:rot-b} implies that the linear case ($p,q=1$) underestimates 
the G-star measurements, while the non-linear relations ($p,q\geq 1$) match the observed 
trend for the more rapid rotators with $P_{\rm rot}\lesssim 5$~d and one 
slower-rotating star (\object{HD 56124}), with $P_{\rm rot}\sim 18$~d. 
Our models reproduce the measurements on some of the stars rather well (e.g. \object{HD 56124}, \object{HN Peg}, \object{$\chi^1$~Ori}, \object{HD 166435}, \object{EK Dra}), but systematically underestimate
the mean field strengths of another group of stars, 
which lie above the model curves for $1<q<2$. 
The offset between the models for $p\sim q=2$ and the observations 
in the range of rotation periods from 20 d to 10 d is roughly 
100--200 G, a substantial difference. 

The G-star sample does not appear to follow any power law with $q>1$
of the form in Eq.~(\ref{eq:Btotcurves}). 
A direct power-law fit to all the G-star data (including the Sun) without 
applying the form of Eq.~(\ref{eq:Btotcurves}) gives 
$q=0.58$, consistent with  the value of $q_{\rm all-obs}=0.58\pm 0.07$ 
found by \citet{Kochukhov20}
for the relation
between $\langle B\rangle$ and the Rossby number. This value is much 
smaller than the other empirical values summarised in 
Sect.~\ref{ssec:obs}. 
The shallow trend, combined with a large scatter, implies only weak 
variation in the mean fields over the entire rotation-period range. 
Within this distribution, the Sun, \object{HD 56124}, and the group 
of stars near $P_{\rm rot}=5$~d fall below the overall pattern. 
The solar-level mean field predicted by this fit is 268~G, much higher 
than expected from SSD fields alone, and approximately only 100~G 
lower than fields measured in K-stars at near-solar rotation rates.

One possible reason for the discrepancy between the model and some 
of the observed stars may be that not all stars in the sample have 
sufficiently similar fundamental parameters. To test this hypothesis, 
we identify the stellar physical parameters that are most likely to 
affect the magnetic activity level or the measurement procedure. Two 
parameters are known to have the strongest influence on the mean 
magnetic field strength at a fixed rotation rate: the effective 
temperature, $T_{\rm eff}$, and the metallicity, represented here by 
the fractional iron abundance
$[\text{Fe/H}] = \log_{10}(N_\text{Fe}/N_\text{H})_{\star} - \log_{10}(N_\text{Fe}/N_\text{H})_{\odot}$ 
relative to solar values. 
\citet{Kochukhov20} adopted a solar iron abundance of -4.58 based on a single Fe~I 5434.5\AA\ line.
However, considering [Fe/H] measurements from various surveys, 
we adopted the standard value of $-4.50$ from \citet{Asplund09}.
These parameters must be taken into account when assessing whether
the mean field depends solely on the rotation rate. 

\begin{figure*}
    \centering
    \includegraphics[width=0.495\linewidth]{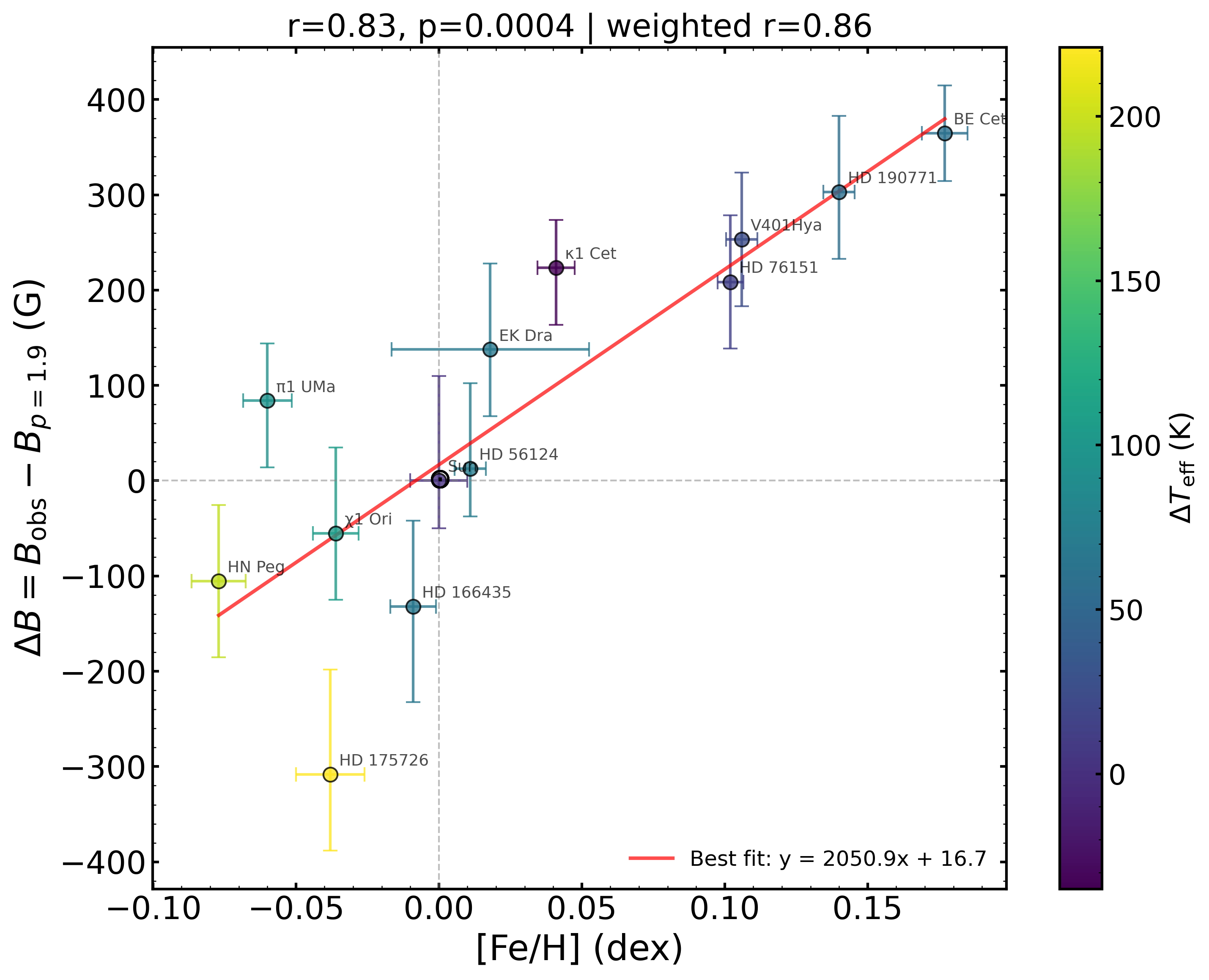}
    \includegraphics[width=0.495\linewidth]{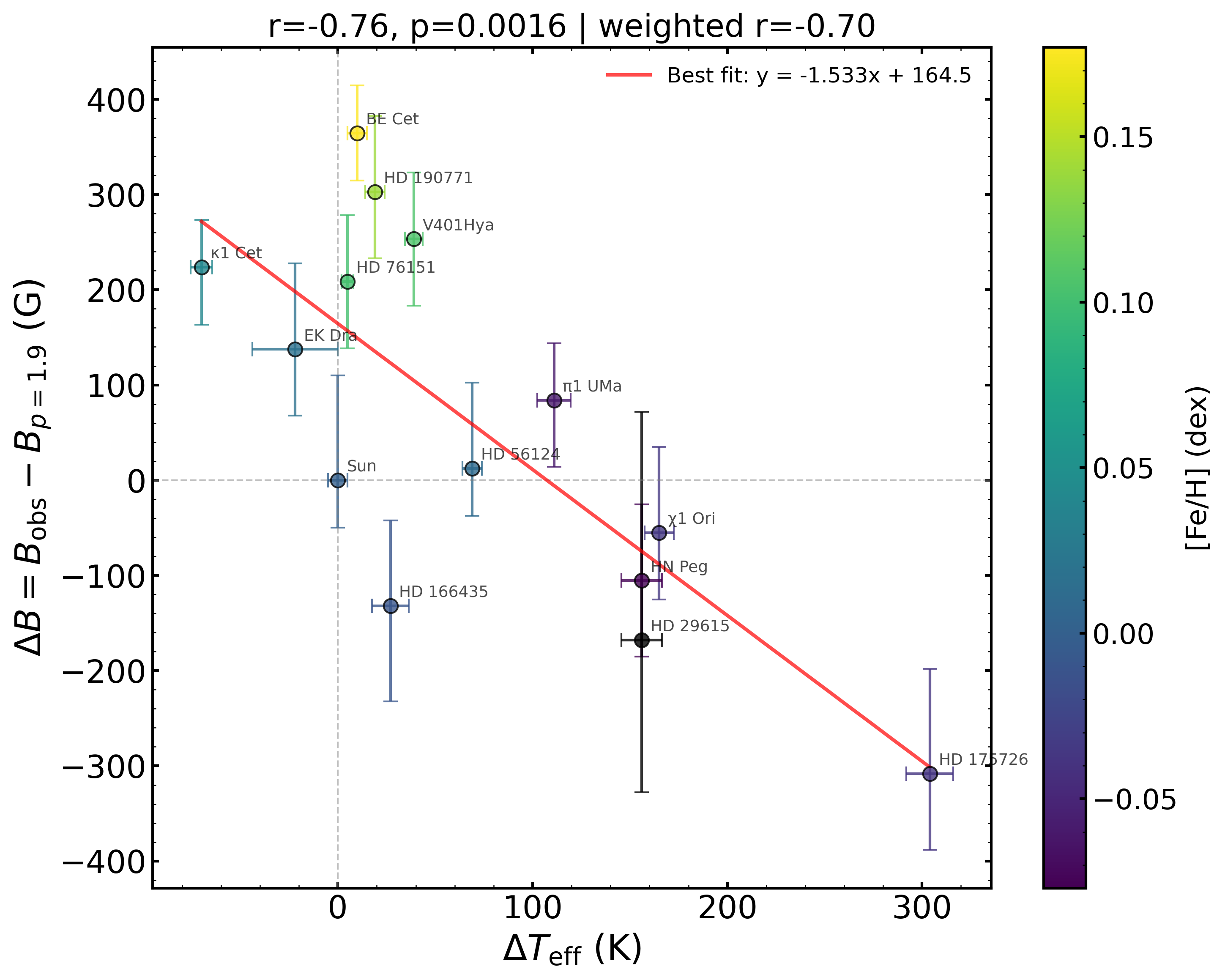}
    \caption{Deviation of the observed mean field strength of the \citet{Kochukhov20} 
    sample from the modelled values with $q=1.9$ in Eq.~(\ref{eq:Btotcurves}), 
    as a function of metallicity (left panel) and the effective temperature difference 
    with respect to the Sun (right panel). The vertical error bars denote the error bounds 
    given by the referred study, and the horizontal ones the standard deviation of the mean 
    [Fe/H] and $T_{\rm eff}$ from the PASTEL catalogue \citep{PASTEL_Soubiran16}. The colour scale shows the temperature (left panel) and metallicity (right panel). The regression lines are shown in red. Vertical lines mark the solar metallicity and reference temperature; horizontal lines show the solar reference field deviation. The correlation coefficient and the $p$ values are given above each frame.} 
    \label{fig:FeH_Teff}
\end{figure*}

\subsection{Metallicity and temperature correlations of model residuals}
\label{ssec:correlation}
The metallicity dependence of stellar magnetic activity (for a fixed Rossby number) 
was already established in various studies \citep{Karoff18,See21,See23,See24}. 
Using literature values, we found that stars along the model curves had near-solar metallicities, while all stars for which our 
model underestimated the mean field had slightly higher metallicities than the Sun, up to 0.20 dex.  
For a few stars the field-strength deviations relative to our model appeared to be additionally affected by differences in effective temperature to the solar value. 

To evaluate the significance of a potential sampling bias, we collected the temperature and 
metallicity measurements for the G-star sample, 
using the PASTEL catalogue of 
stellar atmospheric parameters \citep{PASTEL_Soubiran16}. 
We confined the analysis to studies published after 1994, which are generally based on 
spectra with a relatively high resolution and signal-to-noise 
(the averages did not change but 
the measurement sample got smaller, when we took 1998 as the cut-off). 
To avoid biases introduced by different ways of estimating errors in the 
metallicities used in different studies, we determined unweighted means over 
$N$ measurements and calculated the standard deviation of the mean, 
$\sigma/\sqrt{N-1}$, as the uncertainty.

Figure~\ref{fig:FeH_Teff} shows the mean field strength deviation, 
\begin{equation}
    \Delta B := \langle B_{\rm obs}\rangle - \langle B_{q=1.9}\rangle,
    \label{eq:deltaB}
\end{equation}
of the observed values \citep{Kochukhov20} from our model with $q=1.9$ (a value chosen to jointly optimise the two correlations described below), 
as a function of the metallicity log-difference [Fe/H] (left panel) and the temperature 
difference $\Delta T_{\rm eff}:=T_{\rm eff}-T_{\rm eff,\odot}$ (right panel) 
with respect to the solar values. 
The deviation from the modelled rotation-magnetic field 
relationship turns out to be significantly correlated with the 
metallicity, yielding a coefficient 
of $0.83$ ($p=4\times 10^{-4}$). When the errors are taken into account, 
the weighted correlation becomes $0.86$. A somewhat less significant 
inverse correlation is found for 
$\Delta T_{\rm eff}$, with a coefficient of $-0.76$  ($p=1.6\times 10^{-3}$). Taking into account the uncertainty measures, the weighted correlation becomes $-0.70$. 
This correlation is in line with the empirical and physical  
expectations in both cases, with higher metallicity and cooler temperatures 
being associated with stronger magnetic activity. 
Furthermore, the two correlations share the same target parameter, 
the magnetic field excess: part of the $\Delta B$ 
scatter in the $T_{\rm eff}$ deviations is related to 
systematic differences in metallicity, whereas the scatter in $\Delta B$  with ${\rm [Fe/H]}<0$ is affected by temperature 
deviations, as visualised by the colour code of  the [Fe/H] values (see right-hand colour-bars in Fig.~\ref{fig:FeH_Teff}). 
In the following, we take a closer look at this issue, to rule out any significant 
bias in the abundances we compiled from PASTEL. 

The value $q=1.9$ was adopted for Eq.~(\ref{eq:deltaB}) because it jointly 
optimises both correlations: a systematic scan over $q \in [1.7, 2.0]$ 
shows that $r(\Delta B, {\rm [Fe/H]})$ peaks near $q \approx 1.95$ and 
remains within $0.02$ of its maximum across $1.90 \leq q \leq 2.00$, while 
$r(\Delta B, \Delta T_{\rm eff})$ peaks near $q \approx 1.80$. Both 
correlations are near their respective maxima simultaneously at $q = 1.9$. 
The regression coefficients $a$ and $b$ are insensitive to the specific 
value of $q$ within the range $1.8$--$2.0$.

\subsection{Comparing metallicity estimates}
\label{ssec:metal}

In this section we consider whether the abundances from PASTEL underlying Fig.~\ref{fig:FeH_Teff} may be affected by the presence of a magnetic field, so that the relationship plotted in particular in the left panel of Fig.~\ref{fig:FeH_Teff} could be biased. To test whether there is such a bias, we compare the PASTEL abundances with ones obtained in studies where the authors specifically included the possible effects of the magnetic field on the employed spectral lines. 

The distribution of [Fe/H] measured from individual Fe I lines with a large range 
of Landé factors was shown by \citet[][see their Fig.~6]{Senavci21} for 
\object{EK Dra} (included in the present 
G-star sample). Even for this very active star with 
$\langle B\rangle=$1.40~kG \citep{Kochukhov20}, 
the scatter in single-line [Fe/H] values show 
a weakly increasing trend with $g_{\rm eff}$, making it barely distinguishable 
from the overall scatter about the mean [Fe/H]. This indicates that the larger 
the sample of lines, the more accurate the mean [Fe/H] becomes, 
with only a marginal effect coming from the magnetic sensitivity. 
Indeed, \citet{Senavci21} find [Fe/H] = 0.03 using 190 Fe I lines with 
$0.5\lesssim g_{\rm eff}\lesssim 2.0$, whereas the single-line 
(with $g_{\rm eff}=-0.01$) estimate by \citet{Kochukhov20} yields [Fe/H] = 0.04. 
The PASTEL average for this star (with only 3 measurements after 1994) is 
0.013, leading to 0.017 after adding the later measurement of \citet{Senavci21} of 0.03. 

Another comparison can be made for \object{$\xi$ Boö A}, for which \citet{Hahlin25} 
followed a Monte-Carlo approach to fit four parameters including 
$\langle B\rangle$ and [Fe/H], using the same iron multiplet as in \citet{Kochukhov20} 
but using all four lines together. They found [Fe/H]$\simeq$-0.175 (average over 
results from two spectrographs) while our PASTEL average is -0.147, differing by 
only +0.028 dex. As the authors pointed out, estimations of [Fe/H] using the 
same multiplet might also be prone to biases specific to taking very similar 
lines (except for their $g_{\rm eff}$). As a conservative test, we 
assumed the estimate by \citet{Hahlin25} as more accurate and applied 
a field-dependent correction to PASTEL metallicities, so that the corrected 
metallicity would be
\begin{equation}
[{\rm Fe/H}]^* = [{\rm Fe/H}]_{\rm PASTEL}-0.05\langle B\rangle_{\rm kG},
\label{eq:mag-contamination}
\end{equation}
where the coefficient in the second term
is determined by the point estimate $-0.038$ for \object{$\xi$ Boö A}'s 
[Fe/H] difference of \citet{Hahlin25} measurement from PASTEL value, at $\langle B\rangle=0.78$~kG. When we apply such a `magnetic 
contamination' correction (Eq.~\ref{eq:mag-contamination}) to all PASTEL [Fe/H] estimates of G stars in Fig.~\ref{fig:FeH_Teff}, the correlation 
coefficient found in Sect.~\ref{ssec:correlation} is reduced from 0.83 to 0.79 ($p=2.2\times 10^{-3}$) and the weighted correlation from 0.86 to 0.82. 
This correlation assumes that PASTEL metallicities 
are all systematically overestimated owing to magnetic fields, and is still significant. 

\begin{figure}
    \centering
    \includegraphics[width=\linewidth]{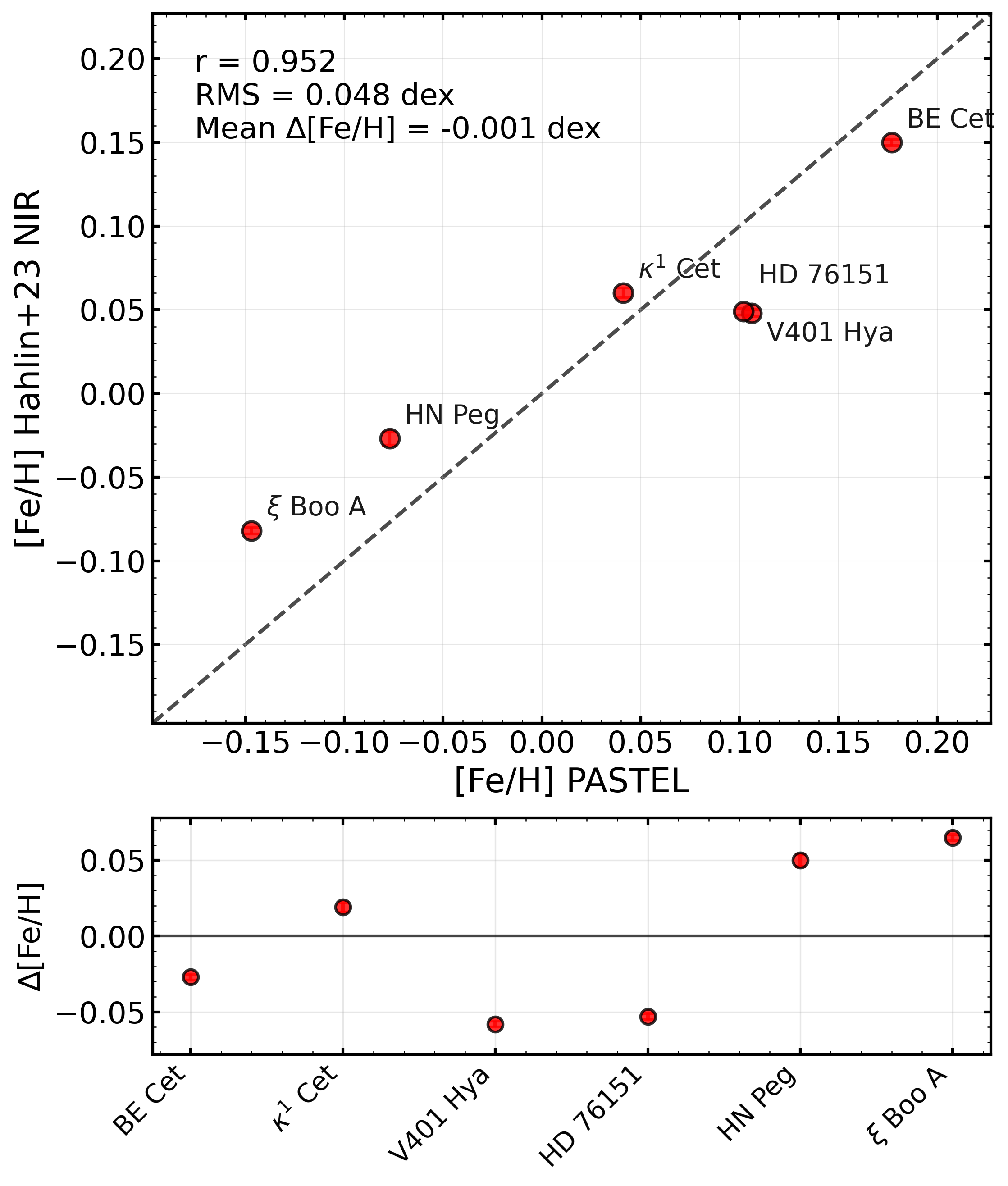}
    \caption{Comparison between [Fe/H] estimates by \citet{Hahlin23} 
    based on six NIR lines with magnetic-field inference and PASTEL 
    averages. The Pearson coefficient, $r$, the root mean square, 
    and the mean 
    differences are indicated. The lower panel shows the individual 
    differences of the NIR estimates from PASTEL values. 
    }
    \label{fig:nir-pastel}
\end{figure}

To assess this potentially systematic error with more measurements, we 
consider the near-infrared spectroscopic analysis carried out by 
\citet{Hahlin23}, focusing on only six stars that were also studied by \citet{Kochukhov20}. For these stars, we compare [Fe/H] values inferred by \citet{Hahlin23} with the 
ones from the PASTEL catalogue. Figure~\ref{fig:nir-pastel} shows the comparison, 
which shows good agreement with a correlation coefficient of 0.95 between 
the NIR-based abundances including the magnetic effects and the standard 
PASTEL abundances. The root mean square difference, ${\rm [Fe/H]_{NIR}} - {\rm [Fe/H]_{PASTEL}}$, 
is +0.048 dex and the mean difference is -0.001 dex. Given all the evidence 
up to this point and the overall agreement in Fig.~\ref{fig:nir-pastel}, 
we rule out any significant systematic effect responsible for the 
correlation shown in Fig.~\ref{fig:FeH_Teff} (left panel) and 
in the following use PASTEL-based abundances, owing to their larger 
statistical samples, without introducing any correction. 

\subsection{Correcting the mean field for metallicity and temperature effects}
\label{ssec:corrections}

The significant correlations we identified in Sect.~\ref{ssec:correlation} 
between model residuals and stellar parameters can be used to 
estimate the effect of the latter on the observed mean field strengths. With that information, this effect can then be removed.  
In this way, we effectively normalised all stars to solar metallicity 
and effective temperature.
Consequently, we isolated the rotation rate as the main dynamo-related 
parameter across the sample of stars considered here, 
allowing us to estimate how emerging flux and the mean field 
strength scale only with rotation. 

We first took the residuals $\Delta B$ in Eq.~(\ref{eq:deltaB}) for each 
star. We then performed a linear multivariate regression of the residuals against
the temperature deviation $\Delta T_{\rm eff}$ and metallicity [Fe/H], 
fitting the model 
\begin{equation}
    \Delta B = a \cdot \Delta T_{\rm eff} + b \cdot {\rm [Fe/H]}
    \label{eq:multiregress}
\end{equation}
without an intercept term, interpreting $\Delta B$ as the predicted 
correction to $B$. This yields correction coefficients of 
$a = +0.300 \pm 0.284$~G~K$^{-1}$ and $b = +2020 \pm 396$~G~dex$^{-1}$ 
with $R^2 = 0.717$. 

\begin{figure}
\includegraphics[width=\columnwidth]{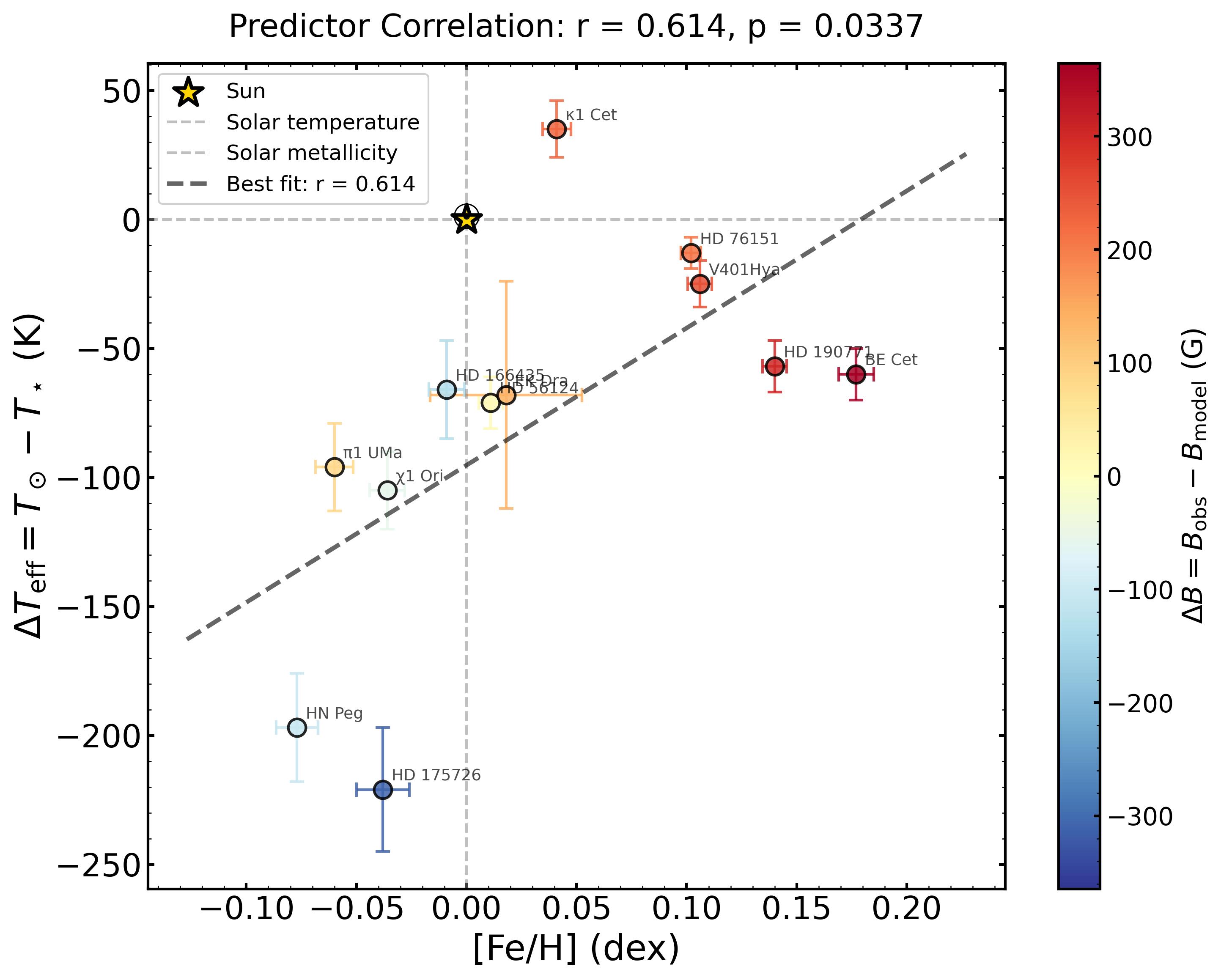}
\caption{Correlation between effective temperature deviation and metallicity 
in our stellar sample ($r = 0.614$, $p = 0.034$). Points are coloured by 
the magnetic field residual $\Delta B = \langle B \rangle_{\rm obs} - \langle B \rangle_{\rm model}$. 
The moderate correlation between predictors creates multicollinearity in the 
multiple regression, inflating coefficient uncertainties. The colour gradient 
shows that both stellar parameters influence $\Delta B$: metal-rich, 
cooler-than-solar stars (upper-right) show the largest positive residuals, 
while metal-poor, hotter stars show negative residuals.}
\label{fig:temp_met_correlation}
\end{figure}

At this point, the relatively large uncertainty on the temperature 
coefficient required an examination of a sampling bias. We checked this by 
considering the predictor correlation, i.e. 
the correlation between the temperature and metallicity differences with 
respect to solar values, $\Delta T_{\rm eff}$ and [Fe/H]. 
Figure~\ref{fig:temp_met_correlation} 
shows that $\Delta T_{\rm eff}$ and [Fe/H] are moderately correlated in 
this particular sample ($r = 0.614$, $p = 0.034$), creating multicollinearity that 
inflates the coefficient uncertainties in Eq.~(\ref{eq:multiregress}), particularly for the temperature. 
The colour-coded 
residuals in the figure demonstrate how both stellar parameters contribute to 
$\Delta B$, as was shown in Fig.~\ref{fig:FeH_Teff}.
While the 
metallicity effect is statistically robust ($\sim$5$\sigma$), the temperature 
coefficient shows lower formal significance ($\sim$1$\sigma$), despite 
$\Delta T_{\rm eff}$ correlating individually with 
$\Delta B$ at $r = 0.76$ ($p = 0.002$; 
Fig.~\ref{fig:FeH_Teff}, right panel). 
We also obtained an optimally weighted parameter defined as a linear 
combination of $\Delta T_{\rm eff}$ and [Fe/H] (Appendix~\ref{sec:app}). 
This parameter gives a correlation of $r = 0.90$ with $\Delta B$, 
supporting that both stellar parameters jointly influence magnetic field 
strength even though their individual contributions are difficult to fully 
disentangle in multiple regression. 

\begin{figure*}
\sidecaption
\includegraphics[width=12cm]{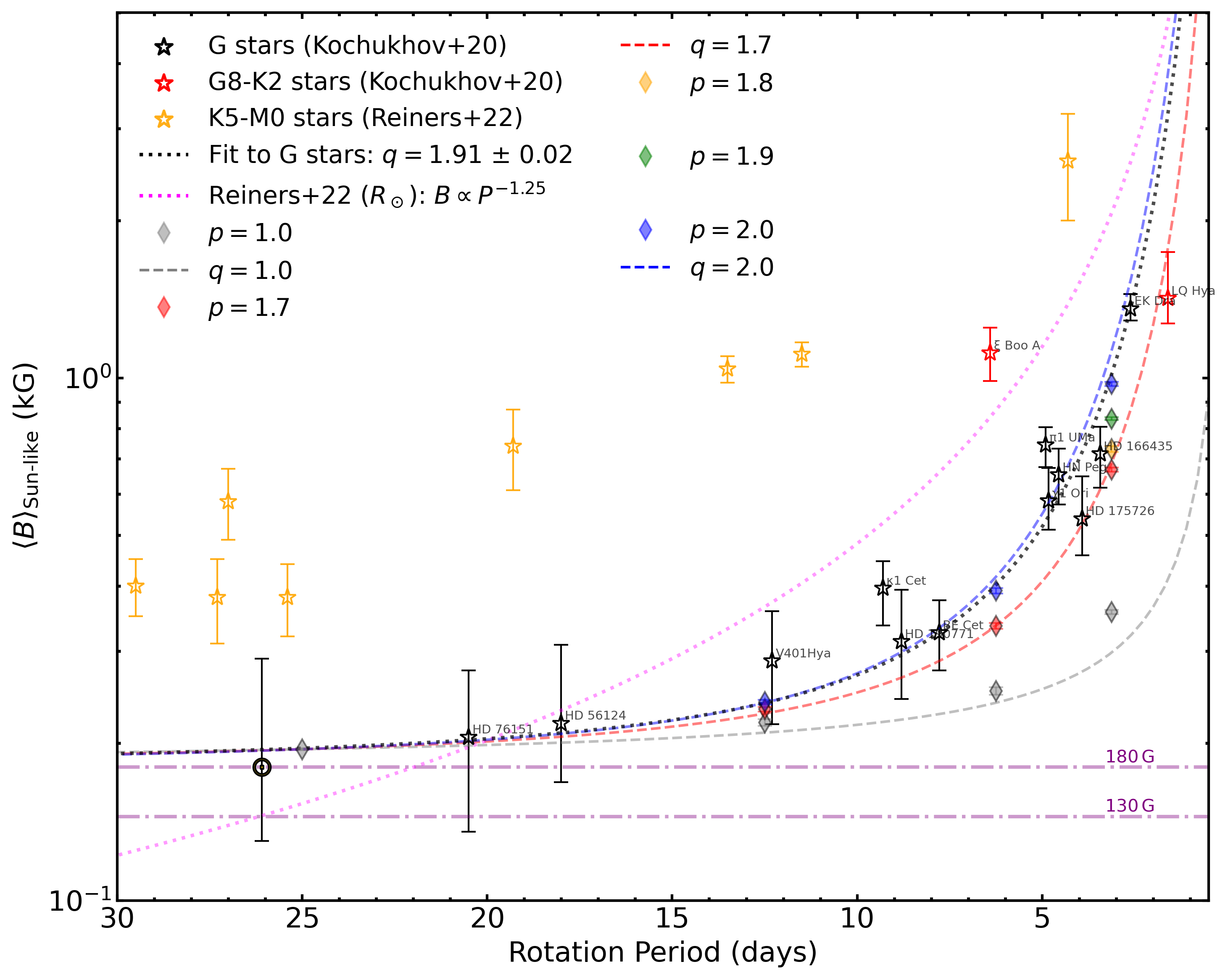}
\caption{Corrected mean magnetic field strength as a function of the 
rotation period following temperature and metallicity corrections 
(Eq.~\ref{eq:Bcorr}). 
See the caption of Fig.~\ref{fig:rot-b} for the definitions. 
}
\label{fig:rot-b-corrected}
\end{figure*}

These correction coefficients indicate that a 100~K temperature decrease 
from solar corresponds to approximately 30~G increase in mean field strength, 
while a +0.1~dex metallicity increase corresponds to approximately 200~G 
increase. 
For context, the mean unsigned surface field of the Sun varies by approximately 10~G between cycle minimum and maximum \citep{Jiang11b,Hofer24}, so the temperature sensitivity amounts to roughly three times the solar cycle amplitude per 100 K difference.
We applied these corrections to normalise all observations to solar 
temperature and metallicity, defining the corrected 
mean field as
\begin{eqnarray}
    \langle B \rangle_{\rm Sun-like} &=& \langle B \rangle_{\rm obs} - a \cdot \Delta T - b \cdot {\rm [Fe/H]}, \\
    a &=& 0.300 \pm 0.284~{\rm G~K}^{-1}, \nonumber \\ 
    b &=& 2020 \pm 396~{\rm G~dex}^{-1}. \nonumber 
    \label{eq:Bcorr}
\end{eqnarray}

The resulting corrected form of Fig.~\ref{fig:rot-b} is shown in 
Fig.~\ref{fig:rot-b-corrected}, where we applied the correction 
only to our G-star sample from \citet{Kochukhov20}. The two significantly cooler stars of 
types G8 and K2 were excluded, as their $T_{\rm eff}$ corrections 
would be likely subject to non-linearities.
Following the correction for 
non-solar stellar parameters, the G-type stars align much closer 
to the model curves (cf. Fig.~\ref{fig:rot-b}), demonstrating how subtle differences in intrinsic 
stellar properties affect field strengths. This may potentially happen through differing degrees of dynamo forcing, but also because line strengths (and hence the amount of Zeeman intensification) depend on metallicity and effective temperature. It also strengthens the possibility that 
globally averaged stellar magnetic fields strengthen in accordance 
with the physical model described in Sect.~\ref{ssec:rot-B} 
(Eq.~\ref{eq:Btotcurves}). 
Also shown is the predictive relationship for the unsaturated branch 
of rotation-magnetism data obtained by \citet[][see their Eq.~2]{Reiners22}, 
where we set the stellar radius to $1R_{\odot}$ to focus on solar-type 
stars. Since the authors took G-star data from \citet{Kochukhov20} results 
as shown in Fig.~\ref{fig:rot-b}, their relationship is possibly affected by 
the metallicity correlation discussed in Sect.~\ref{ssec:correlation}, 
though the effect could be partly corrected by a metallicity-dependent 
variance in stellar radii.

We emphasise that the current correction should be viewed as a first-order adjustment, limited by the small number of G-type stars with reliable magnetic field measurements. 
The moderate correlation between $\Delta T$ and [Fe/H] leads to multicollinearity, amplifying the formal uncertainties of the regression coefficients. 
The metallicity dependence is statistically robust ($5\sigma$), while the temperature dependence should be regarded as tentative ($1\sigma$) and interpreted cautiously.
A larger, homogeneous sample of stars is required to disentangle these effects more reliably.
Thus, the corrected field strengths should be interpreted as first-order adjustments appropriate for early-G stars but they should not be considered as accurate metallicity or temperature dependence outside the restricted parameter range of our sample.

At solar rotation the AR contribution at maximum is only about 8\% of
the total field, implying a cycle-amplitude scatter of about 10--15~G
\citep{Jiang11b,Hofer24}, far below the metallicity corrections of
$\sim$200~G per 0.1~dex applied to the most deviant stars. Multi-epoch
Zeeman intensification measurements of G-type stars spanning our full
rotation range find no statistically significant long-term variation in
$\langle B\rangle$ over baselines of 5--10~years \citep{Kochukhov20},
constraining any cycle-phase scatter to below the per-epoch measurement
uncertainties even where the AR fraction of the total field
exceeds 50\%. Cycle-phase effects therefore remain a secondary contribution
to the scatter throughout the sample \citep[see also][]{Garg25}.

\subsection{Fraction of large-scale fields}
\label{ssec:partition}

In our models, the fraction of the AR-driven fields in the total mean 
field grows with the rotation rate, more or less steeply depending on the 
flux-injection scaling index, $p$. 
This fraction is only 8\% for the solar maximum (though this is likely underestimated owing to the relatively strong mean solar field of 180~G), 
and it reaches up to 82\% for 
$\omega=8$ and $p=2.0$.
Table~\ref{tab:fractional-B_AR} lists this fraction, 
$\langle B_{\rm SFT}\rangle/\langle B_{\rm tot}\rangle$, for different 
rotation rates and flux-injection steepness ($p$) values (see the first five columns). 
This increasing fractional contribution demonstrates that AR fields transition 
from being a minor perturbation to the dominant magnetic flux component as the 
rotation rate increases, while the SSD contribution remains constant. 
The absolute increase in $\langle B_{\rm SFT}\rangle$ from about 15~G at 
solar rotation to about 800~G at $\omega=8$ and $p=2$ directly reflects 
the enhanced flux emergence frequency on faster-rotating stars, which is 
the main driver of the steep rotation-magnetic field relationship we observe.

\begin{table*}[]
    \centering
    \caption{Active-region-driven fraction, emergent and lost flux in models.}    
    \begin{tabular}{cccccccc}
    \hline\hline
        $\omega$ & $P_{\rm rot}$ & $p$ & $\Phi_{\rm AR}$ & $\langle B_{\rm AR}\rangle/\langle B_{\rm tot}\rangle$ & $\Phi_{\rm em}$ & $\Delta\Phi_{\rm tor}/\Delta t$ & $\Delta\Phi_{\rm tor}/\Phi_{\rm em}$ \\
            & [d] &     & [$10^{24}$~Mx] & \% & [$10^{24}$~Mx~yr$^{-1}$] & $10^{24}$~Mx~yr$^{-1}$ & 
            [$\times 10^{-3}$] \\
        \hline
         1 & 25.0 & 1.0 & 0.90 & 7.6 & 4.57 & 0.04 & 7.75 \\
         2 & 12.5 & 1.0 & 2.38 & 18 & 17.2 & 0.15 & 8.96 \\
         4 & 6.25 & 1.0 & 4.36 & 29 & 37.2 & 0.35 & 9.48 \\
         8 & 3.12 & 1.0 & 10.8 & 50 & 100 & 0.98 & 9.73 \\
        \hline 
         2 & 12.5 & 1.7 & 3.19 & 23 & 31.3 & 0.30 & 9.59 \\
         4 & 6.25 & 1.7 & 9.47 & 46 & 91.9 & 0.91 & 9.92 \\
         8 & 3.12 & 1.7 & 29.7 & 73 & 311 & 3.15 & 10.15 \\
        \hline 
         8 & 3.12 & 1.8 & 33.5 & 75 & 388 & 4.36 & 11.22 \\
        \hline 
         8 & 3.12 & 1.9 & 39.9 & 79 & 479 & 6.20 & 12.94 \\
        \hline 
         2 & 12.5 & 2.0 & 3.63 & 25 & 37.2 & 0.36 & 9.60 \\
         4 & 6.25 & 2.0 & 12.9 & 54 & 145 & 1.78 & 12.27 \\
         8 & 3.12 & 2.0 & 48.3 & 82 & 588 & 7.29 & 12.41 \\
    \hline 
    \end{tabular}
    \label{tab:fractional-B_AR}
\end{table*}

\begin{figure}
    \centering
    \includegraphics[width=\linewidth]{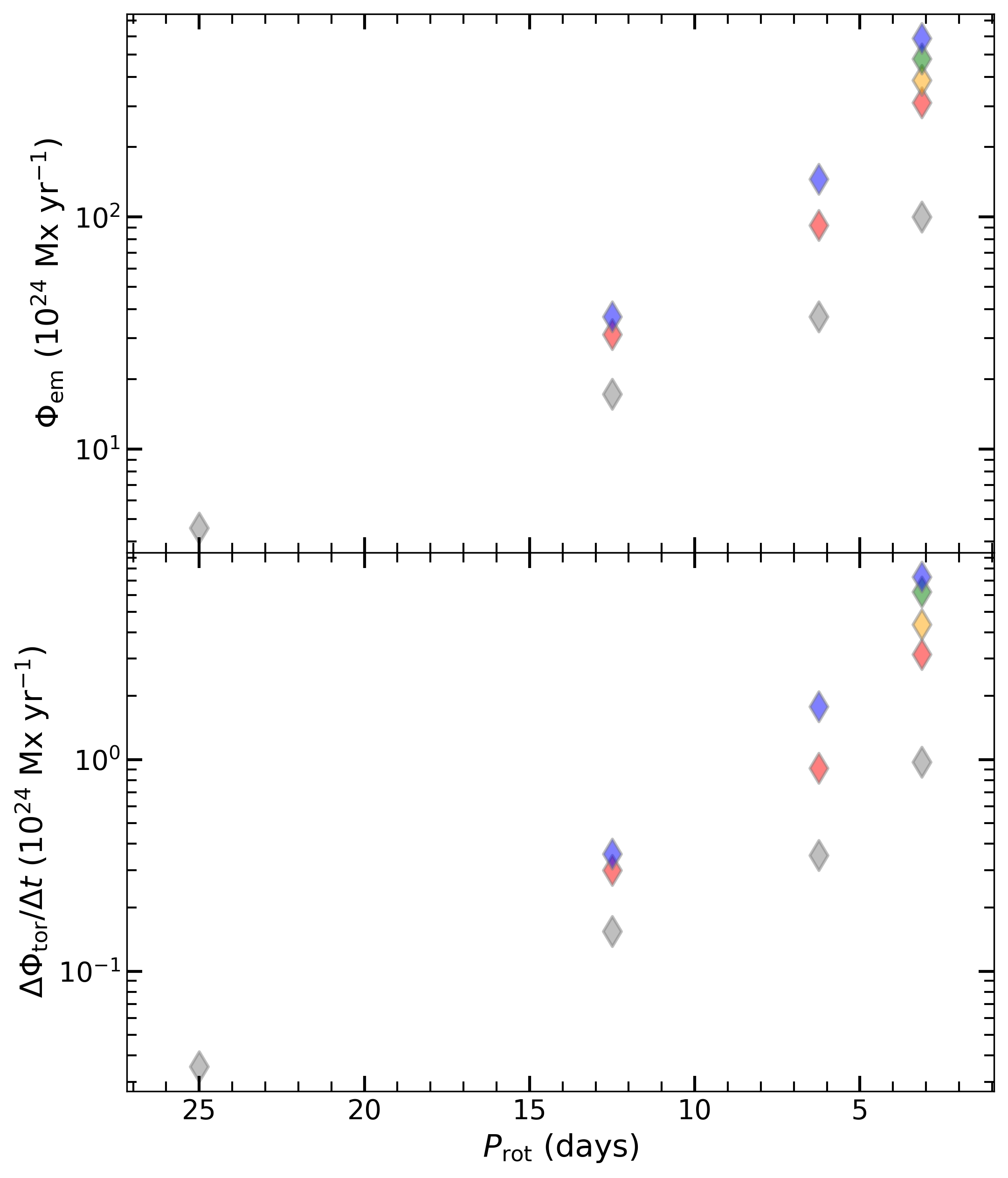}
    \caption{Rate of emerging flux via BMRs (upper panel) and the associated toroidal flux loss rate (lower panel) for simulated cases in Fig.~\ref{fig:rot-b}, using the same symbol colours for $p=1$ to $p=2$.}
    \label{fig:em-loss}
\end{figure}

\subsection{Emerging flux versus toroidal flux loss}
\label{ssec:emloss}
Here, we investigate the AR flux emergence 
frequency as a function of the rotation rate. 
We calculate the rate at which emerging BMRs remove 
toroidal magnetic flux from the stellar interior, 
from slow to fast rotators. 

We calculated the unsigned magnetic flux injected 
by a given emerging BMR, $\phi_{\rm BMR}$, using 
the analytical expression given by \citet{Hofer24},
\begin{equation}
    \Delta\beta = 8.5\times 10^{-10}\sqrt{\frac{\phi_{\rm BMR}}{B_{\rm amp}}},
    \label{eq:phi-BMR}
\end{equation}
where $B_{\rm amp}=374$~G (see Sect.~\ref{ssec:rot-B}), $\Delta\beta$ 
is the angular separation between the BMR polarities. 
Using Eq.~(\ref{eq:phi-BMR}) and summing over 
100 days around the activity maximum, we linearly 
scale the emerging flux to a year, to obtain the 
emerging-flux metric in units of maxwells per year, 
$\Phi_{\rm em}$. 
This quantity expresses the toroidal magnetic flux 
that is lost through emergence at the surface, during the 
activity maximum of a given star.

When flux emerges in the form of an 
$\Omega$-loop, the emerged portion removes a fraction 
of the azimuthally averaged toroidal flux. This fraction is determined 
by the horizontal arc length between the two
polarities, $D=(R_{\odot}\cos\lambda)\Delta\varphi$, which is proportional to 
their longitudinal separation, $\Delta\varphi$ \citep{Cameron20}. 
The polarity separation depends on 
the BMR size (flux), emergence latitude $\lambda$, 
and tilt angle, 
so the relationship between the rotation rate and the amount of
toroidal flux removed from the stellar interior is non-linear. The corresponding toroidal 
flux loss rate can be written as 
\begin{eqnarray}
    \frac{\Delta\Phi_{\rm tor}}{\Delta t} &=& 
    \frac{D}{4\pi R_{\odot}\cos\lambda}\phi_{\rm BMR} \nonumber \\ \nonumber \\
    &=& 
    \frac{\Delta\varphi}{4\pi}\phi_{\rm BMR},
    \label{eq:torloss}
\end{eqnarray}
where $\Delta\varphi$ is determined by preserving the 
inter-polarity distance with increasing latitude. 
Note that the factor 1/4 includes the factor 1/2 
coming from halving the BMR flux to obtain the 
azimuthal flux. In this way, only one of the two opposite-polarity 
footpoints of the flux loop is counted 
when calculating the toroidal flux. 

Figure~\ref{fig:em-loss} shows the rotation-dependence 
of the emerging flux, $\Phi_{\rm em}$, by adding up 
the BMR contributions in Eq.~(\ref{eq:phi-BMR}) 
and the toroidal flux loss in Eq.~(\ref{eq:torloss}), 
both in maxwells per year. The last three columns of 
Table~\ref{tab:fractional-B_AR} list the emerging 
flux rate, the toroidal flux loss, and the ratio 
between the two. It is interesting that the 
fraction of toroidal flux loss from the emerging 
flux stays at the same order of magnitude ($10^{-3}$), 
though gradually increasing with both the rotation 
rate and the rotation-activity steepness, $p$. 

The two panels of Fig.~\ref{fig:em-loss} are not simply proportional to
each other, despite the flux--separation relation in
Eq.~(\ref{eq:phi-BMR}).  The emerging flux sums $\phi_{\rm BMR}$, whereas
the toroidal loss sums $\phi_{\rm BMR}\,\Delta\varphi \sim
\phi_{\rm BMR}^{3/2}$, since $\Delta\beta \propto \phi_{\rm BMR}^{1/2}$
and $\Delta\varphi \approx \Delta\beta\sin\alpha/\cos\lambda$, 
where $\alpha$ is the BMR tilt angle. The latter
relation is only approximate, as the prefactor $\sin\alpha/\cos\lambda$
depends on emergence latitude and the tilt angle, which is itself
a non-monotonic function of both latitude and rotation rate, arising 
mainly from
the Coriolis effect on rising flux tubes \citep{Isik18}.  Larger BMRs
therefore contribute disproportionately more to the toroidal loss than to
the emerging flux, and this non-proportionality has an additional
rotation-rate dependence through the tilt angle, $\alpha$.  
Both effects together cause
the ratio $\Delta\Phi_{\rm tor}/\Phi_{\rm em}$ to increase gradually with
the rotation rate, as listed in Table~\ref{tab:fractional-B_AR}.

\subsection{Comparison with spectropolarimetric data}
\label{ssec:zdi}

\begin{figure}
    \centering
    \includegraphics[width=\linewidth]{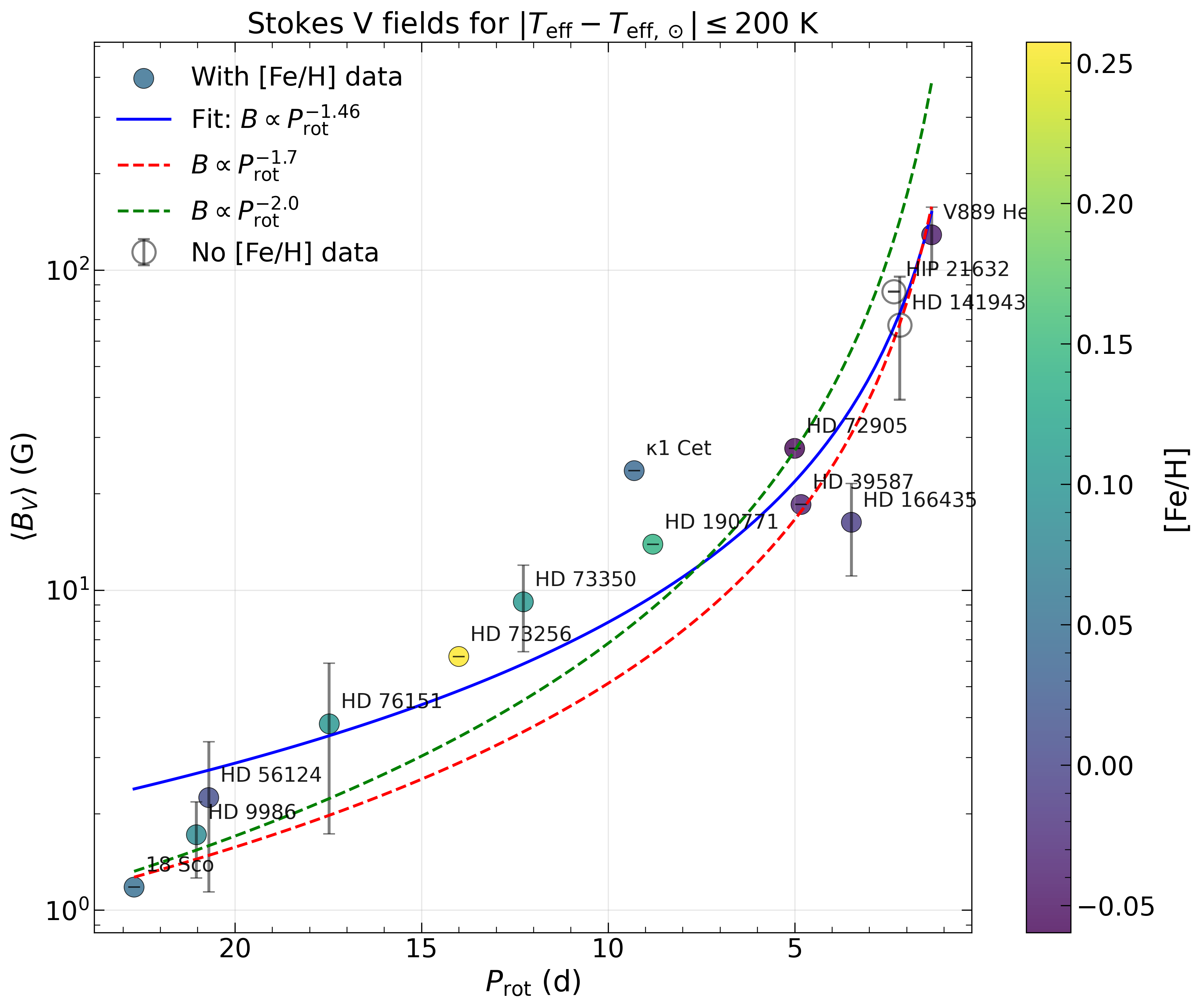}
    \caption{Variation in the mean magnetic field 
    strength inferred from Stokes V as a function of the rotation 
    rate, for stars with the indicated temperature slice taken 
    from \citet{See25}. [Fe/H] averages are taken from the PASTEL 
    catalogue \citep{PASTEL_Soubiran16}. Blue line is a fit to 
    the data, while the dashed lines show our direct scaling of the 
    solar field (here fixed at $\langle B\rangle=1$~G) with rotation. }
    \label{fig:StokesV}
\end{figure}

Following the analysis of data measured using the Zeeman 
effect, we now compare the overall rotation-dependences of our 
models with spectropolarimetric observations of stars with 
near-solar properties but different rotation rates. These measurements 
use the Stokes V component of circularly polarised light, whereby 
opposite-polarity flux elements in the vicinity of each other 
effectively cancel each other's signal, leading to an 
underestimation 
of total magnetic flux \citep{Lehmann19,See19}. Nevertheless, the method can in principle 
capture the large-scale geometry of the field, as well as the 
contribution of surface magnetic features that are segregated 
in the velocity space. 

Figure~\ref{fig:StokesV} shows Stokes V 
measurements of the mean line-of-sight (LOS) magnetic field strength 
from the recent compilation by \citet{See25}. Here, we selected stars 
with a temperature 
within $\pm 200$~K of the solar value of 5777 K. Fitting a power 
law dependence with $\omega^{q_{\rm obs}}$, we find $q_{\rm obs}=1.46$. 
We then implement our scaling laws for $q=1.7$ and $q=2.0$, 
fixing them to $\langle B_V\rangle=1.0$~G at $P_{\rm rot}=25$~d, to 
qualitatively comply with the slowest-rotator measurements in the sample, the closest one to the Sun being \object{18 Sco}. 
Though some of the stars are different than in our analysis with 
Zeeman measurements and $\langle B_V\rangle$ is certainly a different 
field proxy than $\langle B\rangle$, the metallicity gradient becomes 
visible again: slightly metal-rich stars again typically lie above our relation for $q=2.0$, while fitting a free power law to all stars leads to 
a flatter relationship ($q_{\rm obs}=1.46$) than what is found in Sect.~\ref{ssec:correlation} (${q\approx}\,1.9$).

\section{Discussion}
\label{sec:discuss}

Understanding the rotation-activity relationship at the 
fundamental level of mean magnetic field strength, rather than 
through indirect activity proxies, remains an underexplored 
problem in stellar magnetism. 
In this study, we addressed this problem through three 
interconnected parts. First, we established consistent mean 
field strengths for stars rotating more rapidly than the Sun by revisiting published measurements and 
accounting for systematic effects associated with metallicity and effective temperature 
differences (Sect.~\ref{ssec:corrections}). 
Second, we tested how well the FEAT 
framework can reproduce the observed rotation-magnetism trend. 
Third, we used this comparison to determine how the emerging 
magnetic flux, $\Phi_{\rm em}$~-- a quantity that is currently not directly 
observable~-- scales with the rotation rate. 

Our study reveals how the total unsigned surface magnetic flux scales 
with rotation for stars more active than the Sun. For stars 
approaching or exceeding the solar Rossby number ($Ro\gtrsim Ro_\odot$), 
different scaling regimes may apply as large-scale dynamos approach 
criticality \citep{Metcalfe25}. 
This slow-rotator regime is not addressed in the present work.

Our approach compares FEAT models with  
recent direct measurements of 
distributed surface magnetic flux in solar-type stars to estimate the 
rotational scaling of both (a) the surface magnetic flux, $\Phi$, 
and (b) the flux emergence rate, $\Phi_{\rm em}$. 
The framework we introduced 
allows one to compare the measured mean magnetic field strengths of 
solar-type 
stars with different metallicities and temperatures through the 
multivariate regression in Eq.~(\ref{eq:multiregress}). 
For stellar parameter differences 
exceeding those shown in Fig.~\ref{fig:FeH_Teff}, this equation 
should be extended to incorporate potential non-linearities in the 
dependences on $T_{\rm eff}$ and [Fe/H].

The present work has two broad future applications:
($i$) modelling magnetic activity proxies associated with any stellar 
atmospheric layer \citep[e.g.][]{Nemec23}; 
($ii$) setting quantitative constraints on the 
rotational scaling of emerging magnetic flux in solar-like large-scale 
dynamo or flux emergence models. 

The main components of the model are the following. (1) The emergence 
latitudes and tilt angles of bipolar regions are calculated by 
numerical simulations for the buoyant rise of magnetic flux tubes 
\citep{Isik24}, so as to include the gradually increasing importance 
of rotational effects on the distributions of surface magnetic flux. 
(2) The surface flux transport (SFT) module accounts for the 
time-evolution of the field under large-scale flows, as newly emerging regions feed the surface with magnetic flux. 
(3) By adding the rotation-dependent large-scale field in the 
previous step on top of a constant small-scale mean field, 
we derive the scaling relationship between the surface-averaged 
total unsigned field and the rotation rate, allowing us 
to find the optimal rotation-magnetism slope and the associated 
emergent-flux scaling that is consistent with observations. (4) 
Although not an integral part of the model, an important outcome of the 
present analysis is demonstrating the critical need for sample 
homogeneity when deriving magnetic field-rotation relationships. 
This can be achieved either by selecting stars with nearly identical 
fundamental parameters (particularly metallicity and effective 
temperature), or by correcting for biases introduced by sample 
inhomogeneity using methods similar to those described in 
Sect.~\ref{ssec:corrections}. 
Before discussing model aspects that are open to possible improvement, 
we consider the consequences of our model results. 

\subsection{Implications for the nature of stellar magnetic activity}
\label{ssec:imply}

\begin{figure}
    \centering
    \includegraphics[width=\linewidth]{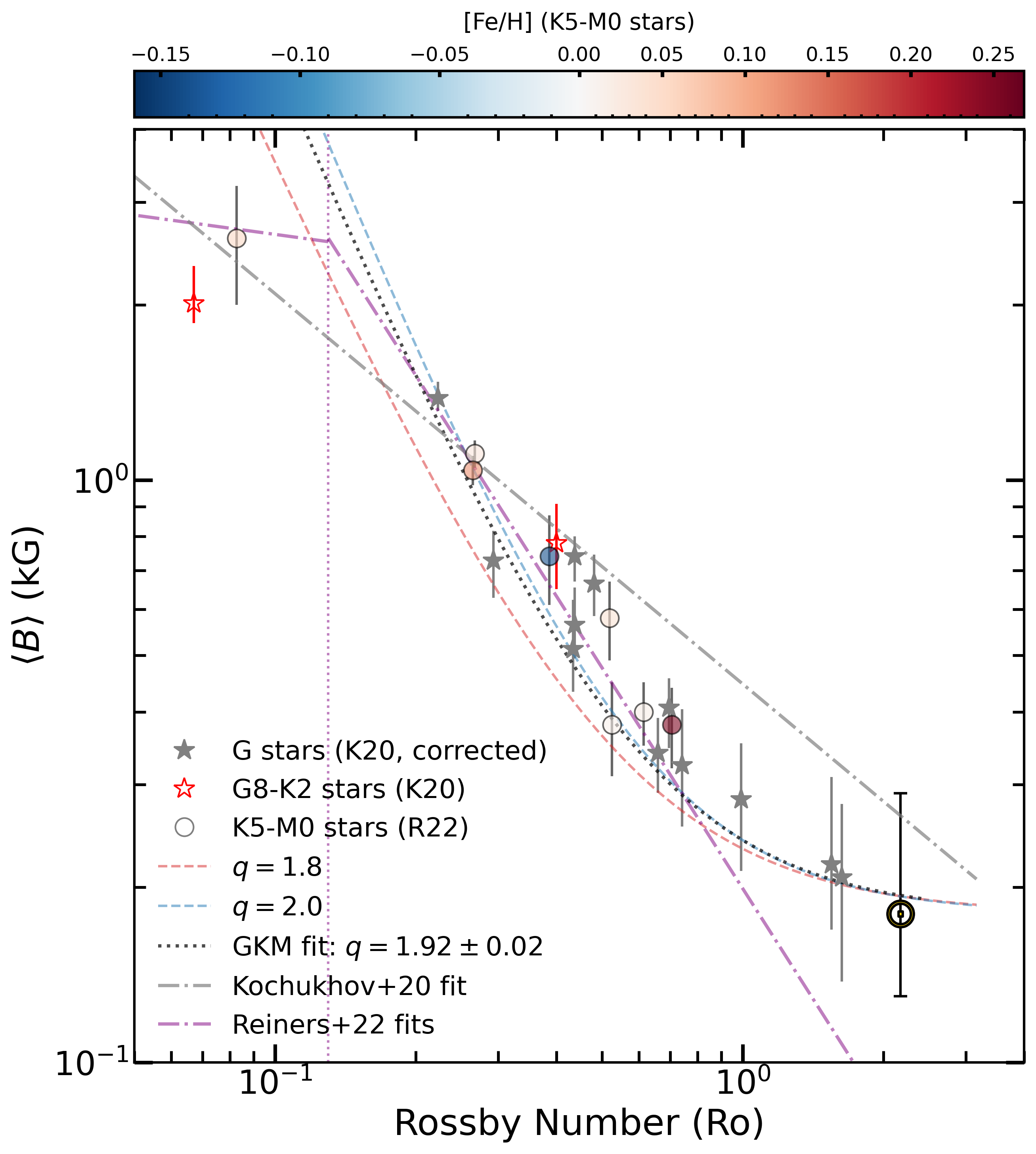}
    \caption{Mean field measurements as a function of the Rossby number, 
    with overlaid model curves, for the same G and K5-M0 stars as in 
    Fig.~\ref{fig:rot-b}. The dotted curve shows a least-squares fit 
    to the G-star subsample as in Fig.~\ref{fig:rot-b}.
    The dash-dotted grey curve 
    shows the power law fit by \citet{Kochukhov20}, including the G8-K2 stars. 
    A vertical dotted line is drawn at $Ro=0.13$, to 
    show the approximate threshold of saturated stellar activity.}
    \label{fig:rossby}
\end{figure}

One important novelty of the present work is the characterisation of 
photospheric sources responsible for Zeeman broadening and intensification 
in solar-type stars of different rotation rates. 
Under the assumption that the global SSD flux is independent 
of the rotation rate and the activity level, we interpret the observed 
rise as the imprint of the increasing frequency of AR emergence, 
so that `network' fields lead to the observed rise of the mean field, on top 
of the constant SSD background. 
In this picture, early-G-type stars with near-solar rotation periods are likely dominated 
by homogeneously distributed, SSD-driven small-scale fields in their 
disc-integrated Zeeman signatures. More than 90\% of the total unsigned 
magnetic flux at any 
time is due to fields generated by a turbulent SSD, i.e. 
formed by magnetoconvection at granular scales. 
With more rapid rotation, the contribution of coherent, 
larger-scale flux elements (BMRs) gradually increases up to 80\% for 
$P_{\rm rot}\sim 3$~d and $p=2.0$, as more and larger ARs are  
presumably sourced from the global dynamo process. 

A cross-comparison of our models with the observed mean fields
from early-G to late-K-type stars becomes feasible, 
when the Rossby number is used 
instead of the rotation period. This is shown in Fig.~\ref{fig:rossby}, 
along with the power-law fit given by \citet{Kochukhov20} for 
all stars and their values in Table~\ref{tab:kochukhov_sample}), 
with an index of 0.67 and the broken power law given by 
\citet{Reiners22} having an index 1.26 for the unsaturated and 
0.11 for the saturated regime. 
For G stars, we applied the empirical correction to $\langle B\rangle$ 
owing to stellar parameter variance (Sect.~\ref{ssec:corrections}),
because in the absence of this correction the major source of scatter 
turns out to be (mainly) metallicity again \citep[see also][]{See24,Pezzotti25,CarvalhoSilva25}. 
The uncorrected version of  
Fig.~\ref{fig:rossby} is given in Fig.~\ref{fig:rossby2}. 
We note that combining stars of different spectral types in such a 
diagram inevitably introduces additional scatter compared to 
more homogeneous samples, as the convective turnover time calibration 
introduces systematic offsets between stellar groups; this is a well-known 
property of rotation--activity relations expressed in terms of the Rossby 
number \citep[e.g.][]{Reiners14}.

Our model is not purely a power law, but a 
combination with a constant term, the background SSD fields 
that we expect are ubiquitous among Sun-like stars of different 
rotation rates. 
Figures~\ref{fig:rot-b-corrected} and \ref{fig:rossby} indicate that 
the model is consistent with the observed values (corrected for 
metallicity and $T_{\rm eff}$ in the case of G stars). We note that the 
regression-based correction procedure was not applied to the sample 
of late-K and early-M stars, because they are too far from solar 
$T_{\rm eff}$ to assume linear dependences to stellar parameters. 
After all, the SSD field is also expected to vary with $T_{\rm eff}$ 
\citep{Bhatia22} and metallicity \citep{Witzke23}.

With the Rossby number decreasing from about 0.8 to 0.4, or the rotation 
period from about 10 d to 3 d for G stars, the fraction of SSD field 
becomes sufficiently small relative to the increasing contribution of 
AR-driven fields, so that the rotation-magnetic field relation approaches 
a power law. Conversely, for ${\rm Ro}\gtrsim 0.8$, 
AR fields 
become gradually less important in determining the disc-integrated 
Zeeman signal, so that 
the rotation-magnetism curve converges towards the predominance of 
the basal flux maintained by SSD action. This regime 
also corresponds to the proposed onset of weakened magnetic braking 
in the late-stage evolution of stellar dynamos \citep{VanSaders16}.

One potential impact of this rotation-magnetism model is expected for 
studies of solar-like stellar dynamos for different rotation rates 
\citep[e.g.][]{Karak14}. Our study provides a first step 
towards constraining how the mean (AR-driven) large-scale field at the 
surface in stellar dynamo should scale with the rotation rate, 
thanks to rigorous comparisons with observations. Furthermore, 
the rate at which the toroidal flux emerges and escapes at the 
surface at a given rotation rate (Fig.~\ref{fig:em-loss}) can be used 
as constraints in narrowing the vast parameter space of 
stellar dynamo models.

\subsection{Systematic effects induced by stellar parameters}
\label{ssec:sys}

The significant correlations we find between model residuals and stellar metallicity as well as temperature (Figs.~\ref{fig:FeH_Teff} 
and~\ref{fig:combined_parameter}) require careful consideration of potential systematic effects in the rotation-magnetism relationship. Five stars in the considered sample (HD 176151, V401 Hya, HD 190771, and BE Cet) all exhibit [Fe/H] > 0 and consistently show magnetic field strengths exceeding our model predictions by over 200 G.

Enhanced metallicity can affect the magnetic field measurements through several mechanisms. Higher metal abundances increase spectral line opacity, potentially enhancing the sensitivity of the transitions to magnetic intensification. In addition, the opacities in the 
bulk of the convection zone would be enhanced, leading to a deeper 
onset of convection at the radiative-zone boundary, effectively 
increasing the dynamo efficiency through the action of a more extended 
convection zone, analogous to a cooler star. The same mechanism 
can be responsible for the temperature anti-correlation in 
Fig.~\ref{fig:FeH_Teff}
that aligns with theoretical expectations that lower-temperature stars 
maintain stronger dynamos at equivalent rotation rates.

Following the stellar parameter corrections presented in 
Sect.~\ref{ssec:corrections}, the best-fit power-law 
slope is consistent with $q_{\rm obs} \approx 1.9$, falling within our modelled range of $1.8$--$2.0$ and serving as a consistency check of
our theoretical framework for stars of near-solar metallicity and temperature. Future 
spectropolarimetric and direct-Zeeman surveys should 
prioritise metallicity- and temperature-controlled samples 
to disentangle the systematic effects from intrinsic 
rotation-activity relationships. Alternatively, for 
early-G-type stars with a $\Delta T_{\rm eff}$ range of $\pm 200$~K 
from the solar value, our model-based corrections found in 
Sect.~\ref{ssec:corrections} can be employed to cover broader 
samples of stars. 

\citet{Pezzotti25} analysed the 
$L_X/L_{\rm bol}$-Ro relationship using FGK stars in the sample given by 
\citep{Wright18}, by splitting the sample in [Fe/H] bins. 
They find that metal-poor stars seem to follow steeper slopes than 
metal-rich stars in the unsaturated regime of the rotation-activity diagram. 
All in all, these results show the importance of treating 
stars of different metallicities separately, and that 
the Sun with its moderate metallicity likely lies on a steeper 
track in the unsaturated regime of the rotation-magnetism relation 
than its metal-rich analogues. 
A qualitatively consistent picture emerges from the chromospheric
activity plane: \citet{CarvalhoSilva25} find that metallicity produces
a clear segregation in the $\log R^\prime_{\rm HK}$--age relation of
near-solar-mass stars, with metallicity-uncorrected chromospheric ages
carrying systematic errors exceeding 3~Gyr.

\subsection{Comparison with Stokes~V-based measurements}
\label{ssec:StokesV}

Recent spectropolarimetric studies based on Stokes V measurements 
have reported scaling trends for large-scale magnetic fields that differ from those inferred from Zeeman-based diagnostics. 
In particular, \citet{Metcalfe25} find that the dipole component inferred from 
ZDI exhibits a concave relationship with the Rossby number, 
rapidly declining to zero near the solar value.  
This behaviour has been interpreted as evidence of a shutdown 
or strong weakening of the global dynamo in slowly rotating stars, potentially linked to the transition to 
weakened magnetic braking 
\citep{VanSaders16}. 

Our analysis 
probes a fundamentally different observable: the surface-averaged total unsigned magnetic field, 
$\langle B \rangle$, 
derived from Zeeman intensification measurements.
This quantity
includes contributions from 
SSD fields and mixed-polarity structures that cancel in Stokes~V but contribute to the total magnetic energy budget. 
Within our framework, a shutdown of the large-scale dynamo 
would 
therefore not imply a vanishing $\langle B \rangle$, but rather a convergence towards a rotation-independent basal field associated with the SSD.
Such a `basal' SSD field, 
$\langle B_\mathrm{SSD} \rangle$, is required in our model to match 
the measured field strengths of slow rotators. 

A direct quantitative comparison between $\langle B\rangle$ and Stokes-V-based field
measurements is non-trivial, as the two
diagnostics respond differently to field geometry, polarity cancellation, and spatial scale \citep{Lehmann19,See19}.
While such a comparison lies
beyond the scope of the present work, we note that metallicity-dependent 
systematics may affect both types of measurements.
As Figure~\ref{fig:StokesV} 
illustrates, relatively metal-rich stars tend to lie above the mean 
$\langle B_V \rangle$ trend, qualitatively similar to the behaviour 
identified for Zeeman-based data (Sect.~\ref{ssec:correlation}). 
This reinforces the conclusion that
sample homogeneity in fundamental stellar parameters is essential for 
deriving unbiased rotation-magnetism relationships, independent of 
the measurement technique employed.

\subsection{Caveats and outlook}
\label{ssec:caveats}

The smooth scaling of the mean magnetic fields 
measured by \citet{Kochukhov20} is broadly consistent with the expectation that 
the AR-producing flux increases at least linearly 
with the rotation rate, or more steeply as we find ($p\sim 1.9$). 
However, the poor sensitivity 
of Zeeman techniques to cool starspot umbrae is an important limitation when 
constraining the scaling exponent, $p$. The ratio of spot to facular 
fields increases with stellar activity level 
\citep{Shapiro+14,Nemec+22}. 
Because magnetic flux in sunspot umbrae, and to a lesser 
extent in penumbrae, is likely underestimated due to their reduced continuum 
intensities, a progressively larger fraction 
of the total flux may be missed in more active stars.
Consequently, the scaling exponent obtained 
here is likely a lower limit to the true scaling exponent. 

Starspot lifetime estimates from photometric surveys \citep{Giles17} 
provide a related observational constraint. We note that FEAT-based 
photospheric-diagnostic simulations \citep[e.g.][]{Nemec23} can derive spot area fractions 
from the magnetic images, and that photometric variability amplitude 
-- used by \citet{Giles17} as a proxy for spot size -- may also 
reflect the degree of AR nesting \citep{Isik20,Karapinar26}. 

In the present study, we employ fixed solar differential rotation and meridional flow profiles (shape and amplitude) for all rotation rates. Observational data suggest that surface differential rotation increases mildly with rotation rate in G-type stars \citep{Balona16}, while meridional flow remains observationally unconstrained for stars other than the Sun. 
Enhanced surface shear would cause stronger flux cancellation 
along elongated polarity inversion lines in active 
regions, leading to weaker surface-averaged fields than predicted by our model \citep{Isik07}. 
Accounting for this effect would therefore require an even steeper scaling of the emerging flux with rotation, again
implying that the inferred exponent $p$ likely exceeds two. 
Both effects are likely second-order in the context of the rotation-magnetism scaling, with the emergence rate being the dominant factor, but their quantitative impact remains to be explored.

We assumed a solar-like cycle for all the stellar rotation rates 
considered, and averaged the mean unsigned field over 100 days 
around the activity maximum. 
The cycle period is fixed as an input parameter, independent of rotation rate, while the initial polar field scaling (Eq.~\ref{eq:Bpol-rot}) serves to maintain internal consistency by preserving the observed correlation between polar field strength and cycle amplitude \citep{Cameron10}. 

While stellar cycle periods are known to be correlated with the rotation period \citep{Brandenburg98}, our simplified approach is appropriate for constraining the emergence rate scaling. First, we measure the quasi-steady state at activity maximum (100-day average), where the surface field reflects the accumulated flux from peak emergence activity rather than being sensitive to cycle period. Second, an SFT parameter study by \citet{Baumann04} as well as $\alpha\Omega$ dynamo-driven SFT \citep{Isik11} showed that while cycle overlap affects temporal modulation, the mean field level at maximum still reflects the integrated emergence rate.

This simplified approach allowed us to focus on 
modelling how the `active state' of the Sun would scale with the 
rotation rate. In reality, the comparison stars we considered might 
well be undergoing different activity levels as they go through 
different phases of their activity cycles, or fluctuations in the 
cases of irregular variability often seen in very active stars. 
Observed or modelled cycle features can be used within our model 
for stars at various cycle phases, which can 
account for part of the variance that remains around 
the rotation-magnetism relationships given in Sect.~\ref{sec:results}.

By scaling the initial polar field with the same exponent as emergence rate (Eq.~2) while maintaining a fixed cycle period, we implicitly assume that dipole-dominated Babcock-Leighton-type dynamos persist across all rotation rates considered. Recent mean-field dynamo calculations by \citet{Zhang24} demonstrate that the dominant magnetic geometry can transition from dipole-dominated to multipolar configurations with increasing rotation rate, with significant implications for differential rotation profiles and meridional circulation patterns. Similarly, 3D global convective dynamo (GCD) simulations by \citet{Viviani18} show complex interactions between different magnetic modes and fundamental changes in convective and rotational influences on field generation at high rotation rates.
The rotation rate beyond which such geometric and symmetry changes become significant for solar-type stars remains poorly constrained.

Recent 3D GCD simulations \citep[][Fig.~1]{Viviani18,Chen25} reproduce observed rotation-magnetism trends after empirical rescaling and exhibit significant cycle-phase variations in $\langle B \rangle$. However, because these simulations do not resolve the near-surface photospheric SSD that contributes the $\sim$180~G rotation-independent baseline in our framework, their fractional cycle amplitude in $\langle B \rangle$ cannot be directly compared with our results: the SSD baseline suppresses the relative cycle modulation in $\langle B \rangle$, particularly at near-solar rotation rates where the AR contribution to the total field is small. Our analysis demonstrates that after correcting for metallicity and temperature dependencies, stars with $10 < P_{\rm rot} < 25$~d and non-solar metallicity and $T_{\rm eff}$ converge towards this $\sim$180~G baseline with steep AR scaling ($\propto \omega^{1.9}$). Whether such simulations would reveal similarly steep emergence scaling when calibrated against systematically corrected observations remains an open question for future comparative studies.

Potential systematic effects can also arise from the determination of stellar metallicities and projected rotation velocities.
\citet{Kochukhov20} derived [Fe/H] and $\varv_{\rm eq}\sin i$ 
solely based on the Fe~I~5434.5~Å line, which has a very small effective Land\'e factor of 
-0.010, and is therefore practically insensitive to Zeeman intensification. 
They estimated that disregarding Zeeman effects in abundance determinations 
could lead to metallicity overestimation of about 0.2~dex for a 1-kG mean field 
when the mean $g_{\rm eff}$ of the lines used is about 1.94. 
This potential error is comparable to the scatter in [Fe/H] 
ensemble averages from PASTEL for the  stars that deviate most strongly from the model predictions (Fig.~\ref{fig:FeH_Teff}, 
left panel).
In addition, thermal effects 
induced by magnetic fields on spectral line formation were not explicitly accounted for in the abundance analysis of \citet{Kochukhov20}. They employed the same 
model atmosphere for both magnetic and non-magnetic regions, arguing that 
temperature differences are primarily captured by the magnetic filling factor 
$f$. However, magnetic fields affect line formation through both 
direct Zeeman splitting and, generally to a larger extent, by modifications of atmospheric thermal structure, 
to which even lines with low Land\'e factors may be sensitive. Solar investigations show that most lines are weakened in the presence of facular magnetic fields \citep{SolankiStenflo84,Solanki93}. 

The magnitude of these systematic errors is likely reduced when metallicity 
estimates are derived from comprehensive analyses using numerous Fe lines 
with a range of $g_{\rm eff}$ values \citep[e.g.][Fig.~6]{Senavci21}, as well as a mixture of Fe I and Fe II lines,
as biases in individual lines would tend to 
average out. This may explain the relatively modest scatter in PASTEL values 
for most stars. For instance, in Sect.~\ref{ssec:metal} we found only a small 
offset between the \citet{Hahlin25} metallicity determination for \object{$\xi$~Boö~A} and 
the PASTEL 
ensemble average. Nevertheless, we cannot rule out that the systematic 
deviations we observe for high-metallicity stars are partially influenced by 
thermal effects on both the $\langle B\rangle$ measurements and the 
conventional abundance determinations. These effects are not fully captured 
by current modelling approaches. 

A natural extension of the present work would be to model 
the large-scale (e.g. dipole) component of stellar magnetic fields in 
SFT simulations and compare the results with Stokes-V-based measurements. Such a 
study would allow one to test whether total unsigned flux and the net signed flux 
exhibit different rotation scalings within the same modelling 
framework, potentially reconciling Zeeman-broadening and 
spectropolarimetric observations \citep{See19}. 

Finally, the present comparison is limited by the small number of solar-type stars with published Zeeman intensification measurements (13 stars in the G-star sample). A larger, more uniformly sampled dataset would allow a more stringent test to be taken of both the rotation--magnetism scaling and the metallicity and temperature corrections derived here.

\section{Conclusions}
\label{sec:conc}
We have modelled the rotation-magnetism relation in solar-type stars using the 
FEAT framework and compared surface-averaged magnetic field strengths with direct 
Zeeman measurements from \citet{Kochukhov20} and \citet{Reiners22}. Our main conclusions are as follows.

\begin{enumerate}
    \item The field measured at the surface scales very similarly to the emerging flux. This is due to the fact that the surface flux transport operates as an effectively linear process.
    \item The observed rotation-magnetism relationship of G-type stars can be physically reproduced by combining a rotation-independent basal flux generated by the SSD with the rotation-dependent global dynamo field driven by flux emergence from the stellar interior. In this scenario, the SSD background dominates for slow rotators, while 
    large-scale flux emergence 
    becomes increasingly important towards rapid rotation.
    \item To match the observational constraints, the magnetic flux emergence rate must scale steeply with stellar rotation, requiring a power-law exponent in the range $p \approx 1.8$--$2.0$.
    This scaling is significantly steeper than the linear dependence assumed in the original FEAT model and implies
    an enhanced dynamo efficiency in faster rotators.
    For the most rapidly rotating stars considered here ($P_{\rm rot}\sim$3~d),
    flux-emergence-driven fields contribute up to 82\% of the total surface magnetic flux including small-scale fields.
    \item Because magnetic fields in dark spots are likely at least partly missed by current measurements and because the fractional contribution of spot fields increases with activity level, the true scaling of the total magnetic flux with rotation may be even steeper than found here.
    \item Systematic deviations between model predictions and observed mean fields for the sample considered here correlate strongly with stellar metallicity ($r=0.83$) and effective temperature ($r=-0.76$), yielding a combined correlation coefficient of 0.90.
    Our framework allows for a first-order comparison of the measured mean fields of stars with different metallicities and temperatures.
\end{enumerate}

These results provide the first quantitative constraints on magnetic flux emergence rates as a function of stellar rotation and establish a foundation for improved stellar dynamo models and magnetic activity diagnostics.

\begin{acknowledgements}
      We thank the anonymous referee, whose comments helped improve the manuscript.
      EI is grateful to T. K{\i}l{\i}\c{c}o{\u{g}}lu (Ankara 
      University) for useful discussions on stellar 
      abundance determinations. 
\end{acknowledgements}

\bibliographystyle{aa} 
\bibliography{aa58620-25} 

\begin{appendix}
\section{Combined correlation of model residuals}
\label{sec:app}

\begin{figure}
    \centering
    \includegraphics[width=\linewidth]{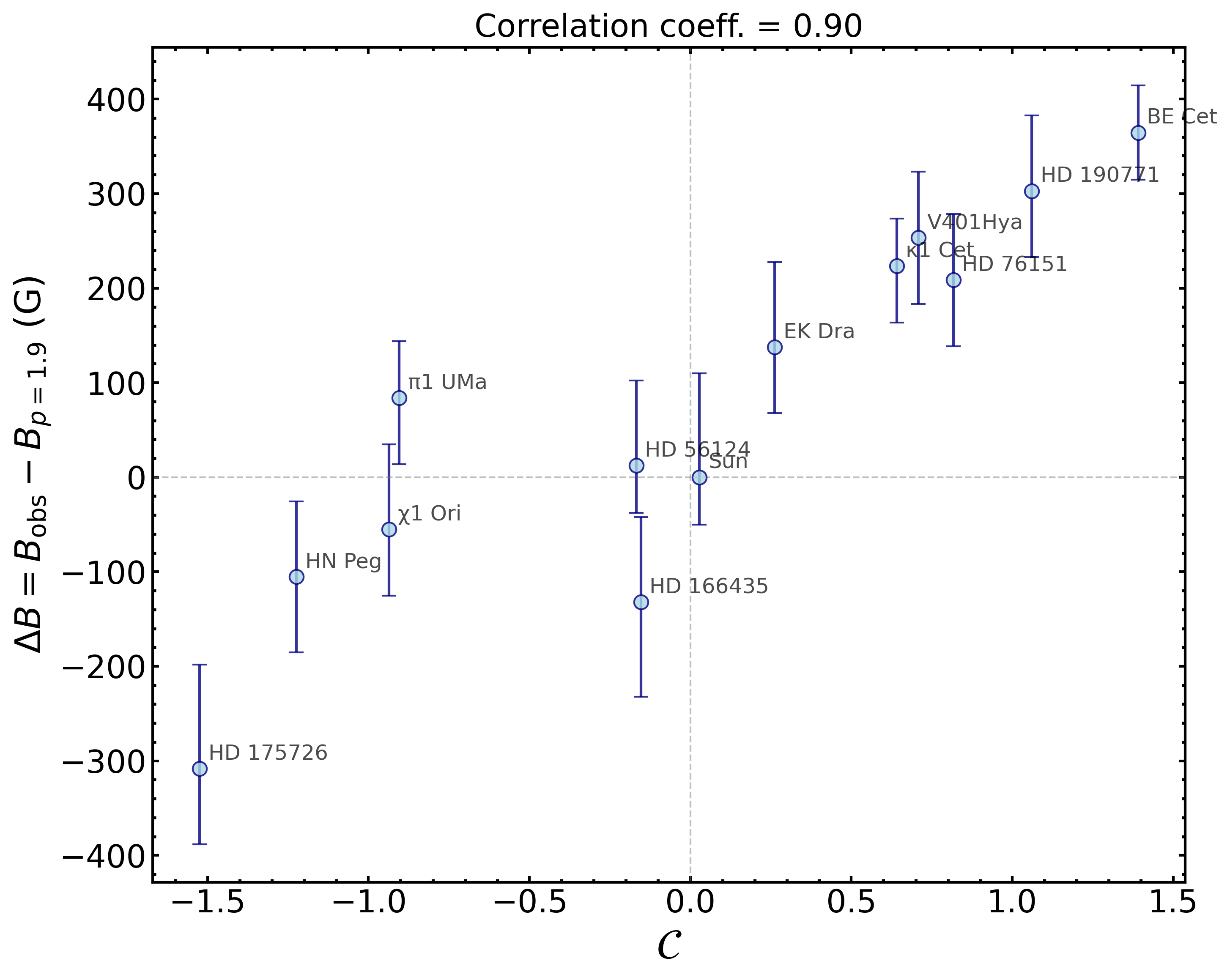}
    \caption{Same as Fig.~\ref{fig:FeH_Teff} but for the combined 
    parameter $\mathcal{C}$ in Eq.~\ref{eq:combined_corr}.}
    \label{fig:combined_parameter}
\end{figure}

The scatter in $\Delta B$ owing to metallicity differences is coupled 
with the one related to temperature differences, as seen from Fig.~\ref{fig:FeH_Teff}. 
We thus introduce a new parameter, by standardising both independent 
variables and forming a weighted linear combination, 
\begin{equation}
\mathcal{C} = -\alpha \frac{\Delta T - \langle\Delta T\rangle}{\sigma_{\Delta T}} + \beta \frac{[\text{Fe/H}] - \langle[\text{Fe/H}]\rangle}{\sigma_{[\text{Fe/H}]}},
\label{eq:combined_corr}
\end{equation}
where $\langle\cdot\rangle$ denotes the sample mean, $\sigma$ the standard deviation, 
and the weights $(\alpha,\beta)$ are optimised to maximise the correlation 
$|{\rm corr}(\mathcal{C},\Delta B)|$, subject to the condition $\alpha+\beta=1$. 
The negative sign 
of the first term in Eq.~(\ref{eq:combined_corr}) is related to the definition 
of $\Delta T$, such that cooler stars (negative $\Delta T$) contribute positively 
to $\mathcal{C}$, aligning with the positive contribution from higher metallicity. 
With the data shown in Fig.~\ref{fig:FeH_Teff}, we find the optimal weights 
$\alpha=0.34$ and $\beta=0.66$, 
yielding a correlation coefficient of ${\rm corr}(\mathcal{C},\Delta B)=0.90$, 
which is significantly higher than for the single-parameter correlations. 
The combined effect of the temperature and metallicity with such weights is thus 
responsible for the high correlation, making these observables the prime suspects 
for the variance in the measured field strengths in stars with similar rotation rates. 
The variation in $\Delta B$ with the combined quantity, $C$, embedding 
the metallicity and temperature effects is shown in 
Fig.~\ref{fig:combined_parameter}. The cluster of stars in the upper-right quarter 
of the plot are those lying significantly above the model curves in Fig.~\ref{fig:rot-b}. 

In agreement with the known physical connections of stellar temperature
and metallicity to the observed activity levels, this result shows that 
the deviation $\Delta B$ of our model from the data is mainly determined 
by these two stellar parameters. In particular, the five stars with 
$\Delta B>200$~G 
(\object{HD 176151}, \object{V401 Hya}, \object{HD 190771}, and \object{BE Cet}) 
are all more metal-rich than the Sun, to the extent that their locations on the 
$(\Delta T_{\rm eff},\Delta B)$ plane reduce the anti-correlation with the 
effective temperature. 
There is still considerable scatter in the combined 
correlation in Fig.~\ref{fig:combined_parameter}. This 
could be related to different phases of stellar cycles 
and different axial inclinations, as well as to measurement uncertainties.

\section{Mean field - Rossby number dependence with uncorrected solar-type stars}
\label{sec:app2}
\begin{figure}
    \centering
    \includegraphics[width=\linewidth]{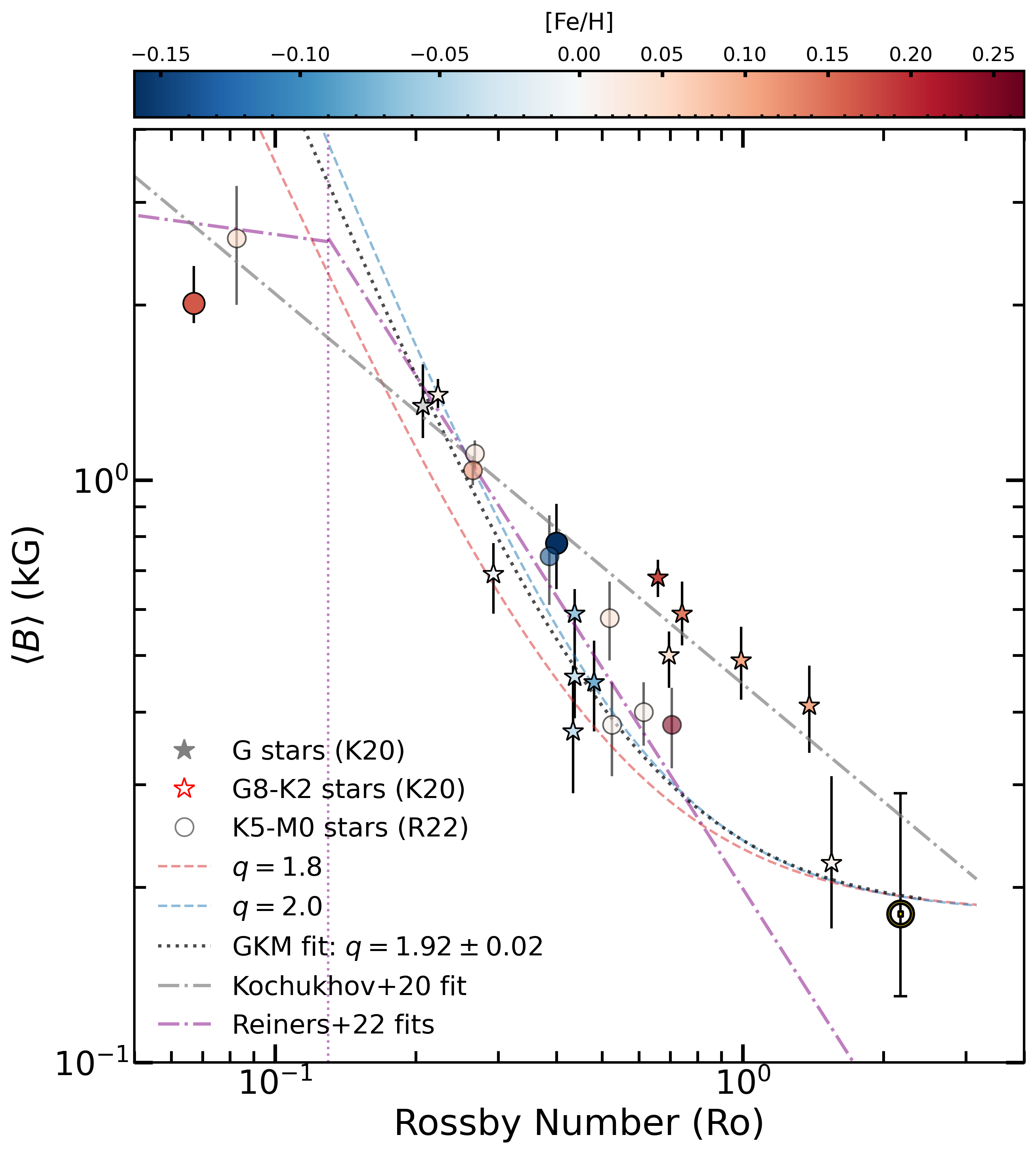}
    \caption{Same as Fig.~\ref{fig:rossby} but taking the measurements 
    of \citet{Kochukhov20} directly, without applying the empirical 
    correction.}
    \label{fig:rossby2}
\end{figure}

Figure~\ref{fig:rossby2} shows the Rossby number 
dependence of the mean field similarly to 
Fig.~\ref{fig:rossby}, but using the uncorrected values 
of the solar-type sample considered in this study. 
When these values are used, the scatter about the 
overall trend increases substantially. 

\end{appendix}

\end{document}